\documentclass{article}
\usepackage{ifpdf}
\usepackage{cite}
\usepackage[pdftex]{graphicx}
\usepackage{array}
\usepackage{amsmath,amssymb,amsthm,mathrsfs}
\usepackage[caption=false,font=normalsize,labelfont=sf,textfont=sf]{subfig}
\usepackage{authblk}
\usepackage{geometry}
\usepackage[font=small,labelfont=md,textfont=it]{caption}
\usepackage{stfloats}
\usepackage[colorlinks,linkcolor=blue,citecolor=blue]{hyperref}
\usepackage{etoolbox}
\usepackage{longtable}
\usepackage{diagbox}
\usepackage{booktabs,makecell,multirow}
\usepackage[capitalise,nosort]{cleveref}
\usepackage{cases,color}
\usepackage{indentfirst}
\usepackage{dsfont}

\crefname{equation}{}{}

\newcommand{\nm}[1]{\lVert {#1} \rVert}
\newcommand{\Nm}[1]{\left\lVert {#1} \right\rVert}

\begin{document}
\title{NFCNN: Toward a Noise Fusion Convolutional Neural Network for Image Denoising
 \thanks
  {
    This work was supported in part by the National Natural Science Foundation of
    China (11771312).
  }}
\author{Maoyuan Xu  \thanks{Email: 663521583@qq.com}
        and Xiaoping Xie  \thanks{Corresponding author. Email: xpxie@scu.edu.cn}\\
        School of Mathematics, Sichuan University, Chengdu 610064, China}

\date{}
\maketitle

\begin{abstract}
Deep learning based methods  have achieved the state-of-the-art performance in  image denoising. In this paper, a deep learning based denoising method is proposed and a module called fusion block is introduced in the convolutional neural network. For this so-called Noise Fusion Convolutional Neural Network (NFCNN), there are two branches in its multi-stage architecture. One branch aims to predict the latent clean image, while the other one predicts the residual image. A fusion block is contained between every two stages by taking the predicted clean image and the predicted residual image as a part of inputs, and it outputs a fused result to the next stage. NFCNN has an attractive texture preserving ability because of the fusion block. To train NFCNN, a stage-wise supervised training strategy is adopted to avoid the vanishing gradient and exploding gradient problems. Experimental results show that NFCNN is able to perform competitive denoising results when compared with some state-of-the-art algorithms.
\end{abstract}

Keywords: Deep Learning, Fusion Block, Two Branches Network Architecture, Image Denoising.

\section{Introduction}\label{section1}
Image denoising is classified as the low level computer vision task which is the basis input of middle and high level tasks, e.g. image segmentation \cite{haralick1985image, ronneberger2015u, wang2018interactive, yang2017suggestive, badrinarayanan2017segnet}, pose estimation \cite{haralick1989pose, erol2007vision, yang2011articulated, cao2018openpose, newell2016stacked, toshev2014deeppose} and image classification \cite{haralick1973textural, lu2007survey, ciregan2012multi, perronnin2010improving, krizhevsky2012imagenet, simonyan2014very}, etc. Since the very first digital image appeared, algorithms for image denoising have been extensively studied aiming to obtain a higher visual quality image with less noise. Nowadays, such methods are divided into two classes in general. One class, also regarded as  model-driven approaches, are  conventional methods based on traditional operations.

  The other class, called data-driven approaches, apply Artificial Neural Networks (ANN) to achieve the state-of-the-art performance in the image denoising field. The conventional methods model noise by employing some mathematical theory to construct the denoising function. Nevertheless, such type methods may suffer from some unexpected effects such as staircase effect, speckle effect and over smoothed effect.
The data-driven algorithms utilize the data-learning ability  of ANN to remove noise from noisy images.

There is an a priori hypothesis for image denoising, i.e. the noise $n$, clean image $x$ and noisy observation $y$ have the following relationship:
\begin{equation*}
  y = x + n,
\end{equation*}
where noise $n$ has lots of modes, such as Additive White Gaussian Noise (AWGN) with standard deviation, Poisson Noise and Salt $\&$ Pepper Noise, etc.
The goal of image denoising is to recover the clean image $x$ from a noisy observation $y$.  A conventional method build a mathematical model to approximate the process $(y-n)$ so as to obtain an approximative latent clean image $\hat{x}$. In recent decades, there have been a considerable number of conventional methods doing well on image denosing, e.g. sparse models\cite{mairal2009non, elad2006image}, nonlocal self-similarity (NSS) models\cite{buades2005non, dabov2007image, xu2015patch} and PDE based models\cite{perona1990scale, chen2000image, bai2007fractional, janev2011fully, xu2021efficient}, etc.

For a data-driven method, thanks to the universal approximation ability of ANN, a function $f$ can be learned from data to obtain, by $\hat{x}=f(y)$, an approximative latent clean image $\hat{x}$ from the noisy observation $y$. There has been significant progress in this field with the development of deep ANNs \cite{mao2016image, zhang2017beyond, zhang2018ffdnet, chen2016trainable, burger2012image, xie2012image, guo2019toward}, especially for deep CNNs \cite{mao2016image, zhang2017beyond, zhang2018ffdnet, chen2016trainable, guo2019toward}.

Inspired by the PM model \cite{perona1990scale}, a Trainable Nonlinear Reaction Diffusion (TNRD) network of good performance was presented in \cite{chen2016trainable} which has a multi-stage structure and applies a stage-wise supervised training strategy during the model training phase.
After TNRD, another awesome network, called DnCNN, was developed in \cite{zhang2017beyond}. Different from TNRD, DnCNN directly creates its model. The architecture of DnCNN is single threaded by stacking convolutional blocks, where one convolutional block, except for the first and last blocks, contains one convolutional layer, one batch normalization layer \cite{ioffe2015batch} and one nonlinear activation layer, and  residual learning \cite{he2016deep} is employed to predict the residual image so as to boost its performance. As shown in \cite{zhang2017beyond}, the denoising performance of DnCNN is enhanced by combining the batch normalization with the residual learning. Later,  a fast and flexible denoising convolutional neural network (FFDNet) was proposed in \cite{zhang2018ffdnet} to denoise images with different noise levels and spatially variant noise, where a noise level map \textbf{\emph{M}} is used to guide the model to obtain a better denoising performance. It should be mentioned that all of the above algorithms are designed for additive white Gaussian noise and their performance is limited for a real-world photograph denoising task. To address this limitation, a convolutional blind denoising network (CBDNet) of outstanding performance was invented in \cite{guo2019toward} by incorporating network architecture, noise modelling, and asymmetric learning. CBDNet contains a noise estimator sub-network to shrink the gap between real-world data and synthetic data, and combines real-world noisy data with synthetic data in the network training so as to make the learned model applicable to real images.

To the best of our knowledge, so far no method fuses noise with the predicted clean image to generate a better result, although some of the above algorithms utilize noise as their input or output. In this work, we propose a Noise Fusion Convolutional Neural Network (NFCNN) to fuse intermediate predicted noise with the intermediate predicted clean image and original input through a fusion block. The information contained in noise of an image is abundant, and noise is not always harmful to the model. It can boost the generalization performance of ANN if the information inside noise is incorporated appropriately.
By the fusion block, NFCNN is able to excavate the information of noise to generate a better denoised result. The fusion block mixes the intermediate predicted clean image, the intermediate predicted noise and the original input to output a fused image. Since the fusion block needs three inputs, our proposed NFCNN needs to simultaneously output the predicted clean image and predicted noise. With these two-branches architecture, a stage-wise supervised training is then taken as our training strategy to boost its generalization performance and to avoid the vanishing gradient and exploding gradient problems. A batch normalization layer is contained in the convolutional block in NFCNN. The branch of NFCNN that outputs predicted noise can be regarded as an application of residual learning. We note that only AWGN is considered to train the model in our work, as it is one of the most universal noise modes.

Heavy experiments demonstrate that our NFCNN trained by the stage-wise supervised training strategy yields competitive denoising performance, when compared with the state-of-the-art methods in terms of PSNR metric, such as BM3D \cite{dabov2007image}, WNNM \cite{gu2014weighted}, TNRD \cite{chen2016trainable}, DnCNN \cite{zhang2017beyond} and FFDNet \cite{zhang2018ffdnet}. NFCNN surpasses FFDNet by 0.02$\thicksim$0.34dB and DnCNN by 0.0$\thicksim$0.17dB on BSD68 dataset. On CBSD68 dataset, when the noise level $\delta$ are 15 and 25, DnCNN outperforms FFDNet by 0.02dB, and NFCNN exceeds DnCNN by 0.01$\thicksim$0.02dB. Another well known dataset for image denoising is Set12. The proposed NFCNN outperforms FFDNet by 0.03$\thicksim$0.04dB when $\delta=$ 50 and 75. DnCNN has better performance than FFDNet when $\delta=15$, and the best performance still belongs to NFCNN. Experiments are also conducted on Kodak24 dataset and McMaster dataset and give  similar results. More detailed training setting and experimental results are described in \cref{section4}.

The rest of this work is organized as follows. \cref{section2} gives a brief summarization of recent related works. \cref{section3} presents a  high level concept of our proposed NFCNN network architecture and a detailed description of the fusion block. Heavy experimental results are displayed in \cref{section4} to demonstrate the denoising performance of our method. Finally, \cref{section5} is devoted to concluding remarks.

\section{Related Work}\label{section2}
As mentioned before, there are mainly two classes of  methods aiming to image denoising: conventional methods based on traditional operations, and data-driven methods based on ANN. Since the proposed method is related to ANN, some of the important works in this field will be summarized in this section.
\subsection{TNRD}
The basic idea of TNRD by Chen \& Pock \cite{chen2016trainable} is to design a learning based nonlinear reaction diffusion model for image processing. Before TNRD, Perona and Malik \cite{perona1990scale} proposed the following classic nonlinear diffusion PDE called PM model for image denoising:
\begin{equation}
\left\{
\begin{aligned}
  & \frac{\partial u}{\partial t} = \text{div}(g(|\nabla u|)\nabla u), \\
  & u|_{t=0}=f, \\
\end{aligned}
\right.
\label{PM_model}
\end{equation}
where $\nabla$ denotes the gradient operator, $\text{div}$ is the divergence operator, $t$ represents the time, and $f$ is an initial image to be processed.   $g(\cdot)$ is regarded as an edge-stopping function\cite{black1997robust} here. A typical $g$-function is of the form $g(z)=1/(1+z^2)$.

\begin{figure}[!t]
  \centering
  \includegraphics[width=5in]{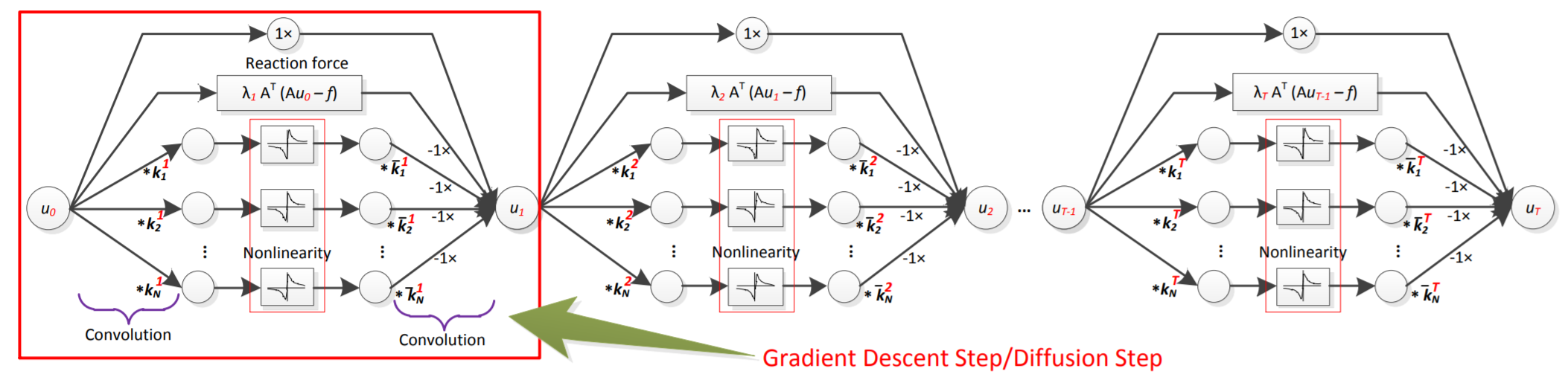}\\
  \caption{The architecture of TNRD.}\label{TNRD_network_architecture}
\end{figure}

Evolved from the discretization scheme of \cref{PM_model}, Chen \& Pock proposed a multi-stage trainable nonlinear reaction diffusion model of the form
\begin{equation*}
  \frac{u_t-u_{t-1}}{\Delta t} = -\underbrace{\sum_{i=1}^{N_k}{K_i^t}^{\top}\phi_i^t(K_i^tu_{t-1})}_{\text{diffusion term}} - \underbrace{\psi^t(u_{t-1}, f)}_{\text{reaction term}},
\end{equation*}
where $K_i\in\mathbb{R}^{N\times N}$ is a highly sparse matrix and $\Delta t$ is set to $1$ in practice. Functions $\phi_i^t$ and $\psi^t$  describe the diffusion and reaction processes, respectively. Matrix $K_i$ can be regarded as a convolutional operator  and can be replaced by a learnable module, i.e. artificial neural networks, and $N_k$ is set to be the number of filters. The concept graph of architecture for TNRD is summarized as in \cref{TNRD_network_architecture}.

Due to its deep network architecture, TNRD is easy to encounter the vanishing gradient issue during the training process. Thus, a stage-wise scheme of greedy training is considered as a trick for training to optimize the cost function
\begin{equation*}
  L(\Theta_t) = \sum_{s=1}^{S}l(u_t^s, u_{gt}^s),
\end{equation*}
where $u_t^s$ is the output of stage $t$, $u_{gt}^s$ is the ground-truth label of stage $t$, and  $l(\cdot,\cdot)$ denotes the $L_2$ loss.

Heavy experiments show that TNRD has  better performance than the conventional methods not only in image denoising, but also in some other aspects like  single image super resolution and JPEG deblocking. There are three possible points to explain why TNRD surpasses other methods in terms of PSNR metric \cite{chen2016trainable}:
\begin{itemize}
  \item \emph{Anisotropy}. Convolutional filters are obtained by training, which may lead to different kernels with anisotropy along different directions.
  \item \emph{Higher order}. The learned filters can have assorted (even fractional) orders of derivatives.
  \item \emph{Adaptive forward/backward diffusion through the learned nonlinear functions}. Nonlinear mappings or functions are also acquired by training  to enrich the smoothing ability of diffusion process.
\end{itemize}
In addition, the trained model is lightweight to run on GPU.

\subsection{DnCNN}
DnCNN \cite{zhang2017beyond} is a method that incorporates residual learning\cite{he2016deep} and batch normalization\cite{ioffe2015batch} so as to achieve good performance in image restoration.
According to the experiments executed with DnCNN, residual learning and batch normalization not only speed up the training process, but also largely boost the generalization performance of DnCNN. Unlike the residual learning that employs a large amount of residual units in network, DnCNN uses a single residual unit to predict the residual image. Batch normalization has the ability to speed up the training phase and enhance performance in the inference process, and can also avoid overfitting to some extent.

\begin{figure}[!t]
  \centering
  \includegraphics[width=5in]{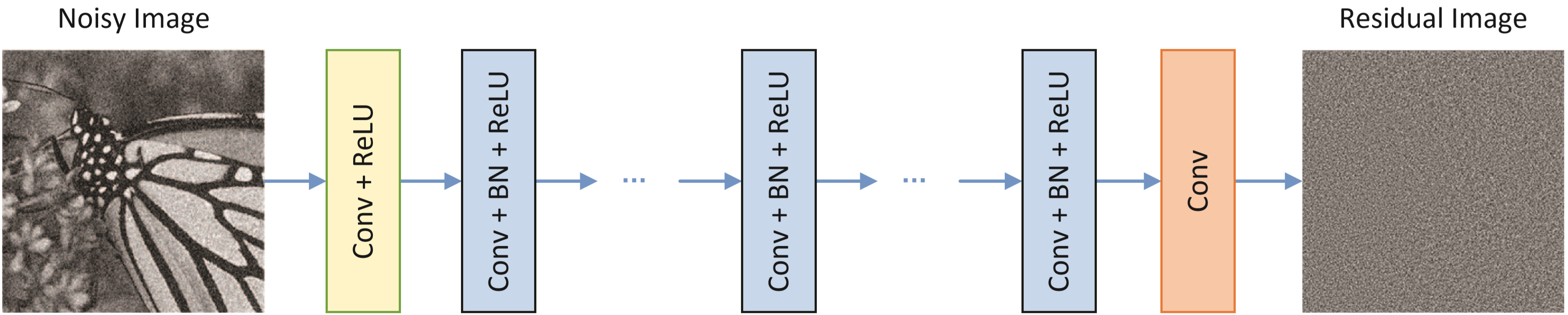}\\
  \caption{High level architecture of DnCNN.}\label{DnCNN_network_architecture}
\end{figure}

It has been shown in the experiments of DnCNN that the training becomes more stable and leads to better testing performance after combining residual learning and batch normalization in the training process. DnCNN employs a $3\times3$ kernel in its convolutional layers, and adopts an appropriate receptive field to match the patch size used in conventional methods. A high level architecture of DnCNN can be described by \cref{DnCNN_network_architecture}. The loss function of DnCNN is
\begin{equation*}
  l(\Theta) = \frac{1}{2N}\sum_{i=1}^{N}\nm{R(y_i;\Theta)-(y_i-x_i)}_{F}^{2},
\end{equation*}
where $\Theta$ is the trainable parameters vector, $N$ represents the number of training samples, $F(y)=x$, and $R(\cdot)$ is the residual mapping.

Depth of DnCNN network is set to be 17 for gray scale image denoising and 20 for color image denoising, respectively. Expensive experiments show that DnCNN can achieve state-of-the-art performance. DnCNN  performs well not only in image denoising, but also in some other fields like single image super-resolution and JPEG image deblocking.

\subsection{FFDNet}
Aiming to handle the cases with different levels of noise and spatially variant noise, FFDNet \cite{zhang2018ffdnet} adopts a non-uniform noise level map \textbf{\emph{M}} as one input. For an original image of size $W\times H\times C$, four downsampled sub-images of size $\frac{W}{2}\times\frac{H}{2}\times4C$ are utilized to improve the efficiency of the network. Here $W$ is the width of the image, $H$  the height, and $C$ the number of channels with $C=1$ for a gray scale image and $C=3$ for a color image.

\begin{figure}[!t]
  \centering
  \includegraphics[width=5in]{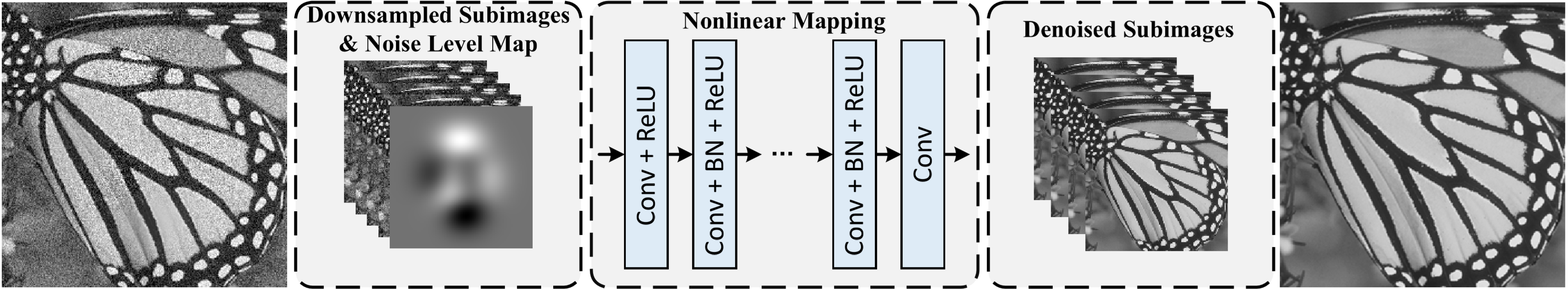}\\
  \caption{The architecture of FFDNet for image denoising.}\label{FFDNet_network_architecture}
\end{figure}

When the level of noise in the testing process has a large gap to the one used in the training process, the result will be over-smooth or under-smooth. With the help of the noise level map \textbf{\emph{M}}, FFDNet is able to deal with the spatially variant noise problem, that is regarded as a challenge in the image denoising field, with  high performance. Similar to DnCNN, FFDNet contains residual learning and batch normalization so as to boost its generalization performance, and the convolutional kernel size  is set to be $3\times3$. The number of convolutional layers is empirically set to be 15 for a gray scale image and 12 for a color image, respectively. It should be noted that FFDNet is not to predict the residual image. The whole architecture of FFDNet can be illustrated by \cref{FFDNet_network_architecture}.

Since FFDNet uses the noise level map \textbf{\emph{M}} as one of inputs and does not predict the residual image, its loss function, different from that of DnCNN, is of the form
\begin{equation*}
  l(\Theta) = \frac{1}{2N}\sum_{i=1}^{N}\nm{F(y_i, M_i;\Theta)-x_i}^{2},
\end{equation*}
where $\Theta$ is the  trainable parameters vector, $N$  the number of training samples, $M_i$  the corresponding noise level map for sample $y_i$, and $F(\cdot, \cdot)$ the nonlinear mapping modelled by FFDNet. Notice that Adam algorithm \cite{kingma2014adam} is employed as an optimizer in the training process of FFDNet to minimize $ l(\Theta)$.

\subsection{CBDNet}
\begin{figure}[!t]
  \centering
  \includegraphics[width=5in]{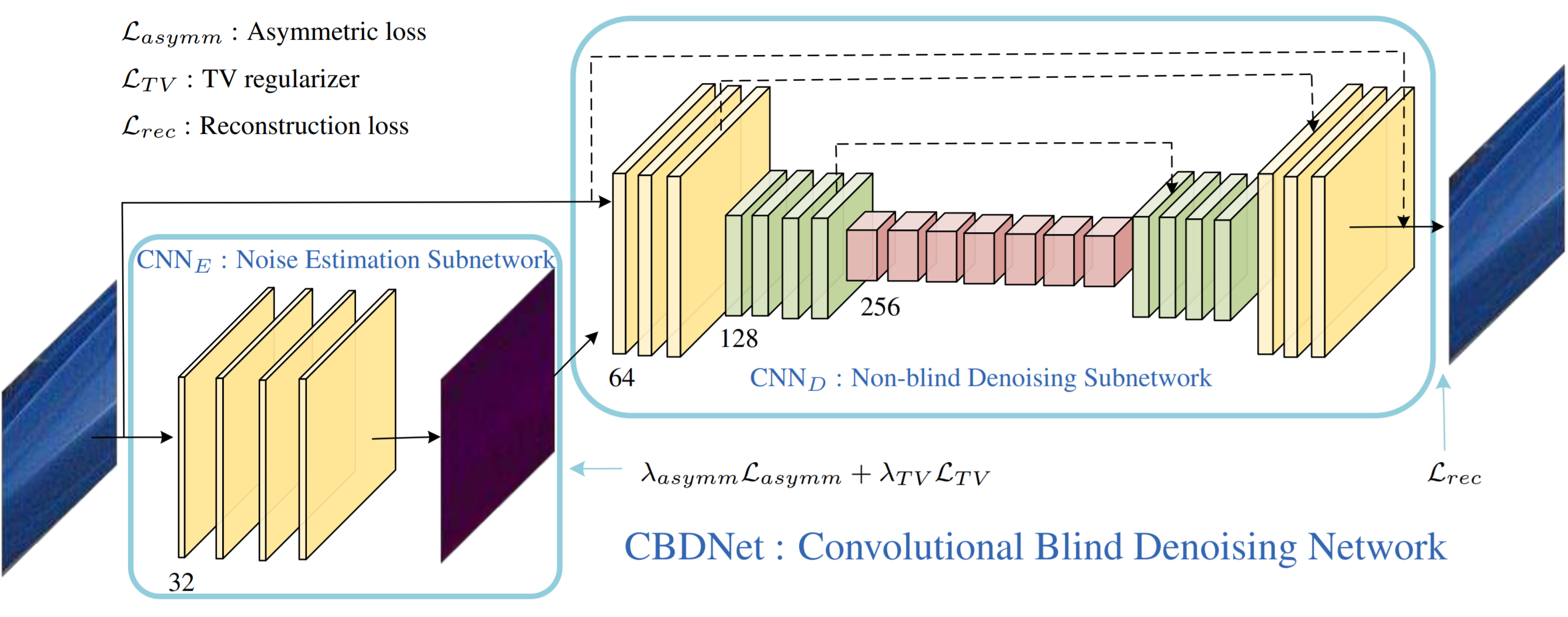}\\
  \caption{Illustration of CBDNet for blind denoising of real-world noisy photograph.}\label{CBDNet_network_architecture}
\end{figure}

Though having achieved impressive performance in removing Additive Gaussian White Noise, deep CNNs are limited on real-world noise. CBDNet \cite{guo2019toward} is produced to improve the robustness and practicability of deep denoising models. On one hand, the denoising performance of CNN largely depends on the noise level gap between the real-world data and synthetic data. On the other hand, the distributions of noise from the real world and synthetic data are actually different, which indicates that training is under two different domains. CBDNet utilizes a subnetwork as the noise estimation extractor to decrease the difference between the two domains, and uses an asymmetric loss to make it robust. Illustration of the CBDNet architecture is shown by \cref{CBDNet_network_architecture}.

The loss function, $L$, of CBDNet consists of three parts, i.e.
\begin{equation*}
  L = L_{rec} + \lambda_{asymm}L_{asymm} + \lambda_{TV}L_{TV},
\end{equation*}
where $L_{rec}, $ $L_{asymm} $ and $L_{TV}$ denote  respectively the reconstruction loss, asymmetric loss and TV loss, and $\lambda_{asymm}$ and $\lambda_{TV}$ are the corresponding weight factors.
The reconstruction loss $L_{rec}$ is defined as
\begin{equation*}
  L_{rec} = \Nm{\hat{\textbf{x}}-\textbf{x}}_2^2,
\end{equation*}
where $\textbf{x}$ represents the output of CBDNet.
According to the estimated noise level $\hat{\sigma}(y_i)$ and the ground-truth noise level $\sigma(y_i)$ at $i$-th pixel, the asymmetric loss $L_{asymm}$ and TV loss $L_{TV}$ can be formulated as
\begin{eqnarray*}
  L_{asymm} &=& \sum_{i}|\alpha-\mathbb{I}_{(\hat{\sigma}(y_i)-\sigma(y_i))<0}|\cdot(\hat{\sigma}(y_i)-\sigma(y_i))^2,\\
  L_{TV} &=& \Nm{\nabla_h\hat{\sigma}(\textbf{y})}_2^2 + \Nm{\nabla_v\hat{\sigma}(\textbf{y})}_2^2,
\end{eqnarray*}
respectively,
where $\mathbb{I}_e=1$ for $e<0$ and $\mathbb{I}_e=0$  for $e\geq 0$,  $\alpha$ is a hyperparameter,
and $\nabla_h$ and $\nabla_v$ are gradient operators along the horizontal and vertical directions.

\section{The Proposed NFCNN Network}\label{section3}
This section is to construct a noise fusion convolutional neural network (NFCNN) for image denoising. NFCNN has a hierarchical structure to build a deep neural network (DNN). As we know, one of the drawbacks of DNN is the difficulty of training. We then take a stage-wise supervised training strategy during the training phase to avert vanishing gradient and exploding gradient issues. The fusion block fuses the intermediate predicted clean image, the intermediate predicted noise and the original input to generate a mixed output. The hint of fusion block comes from the fact that the exchange of information in network is always helpful for deep learning. The output after the fusion block will be taken as an input of next stage.

\subsection{Network Architecture}
\begin{figure}[!t]
  \centering
  \includegraphics[width=4in]{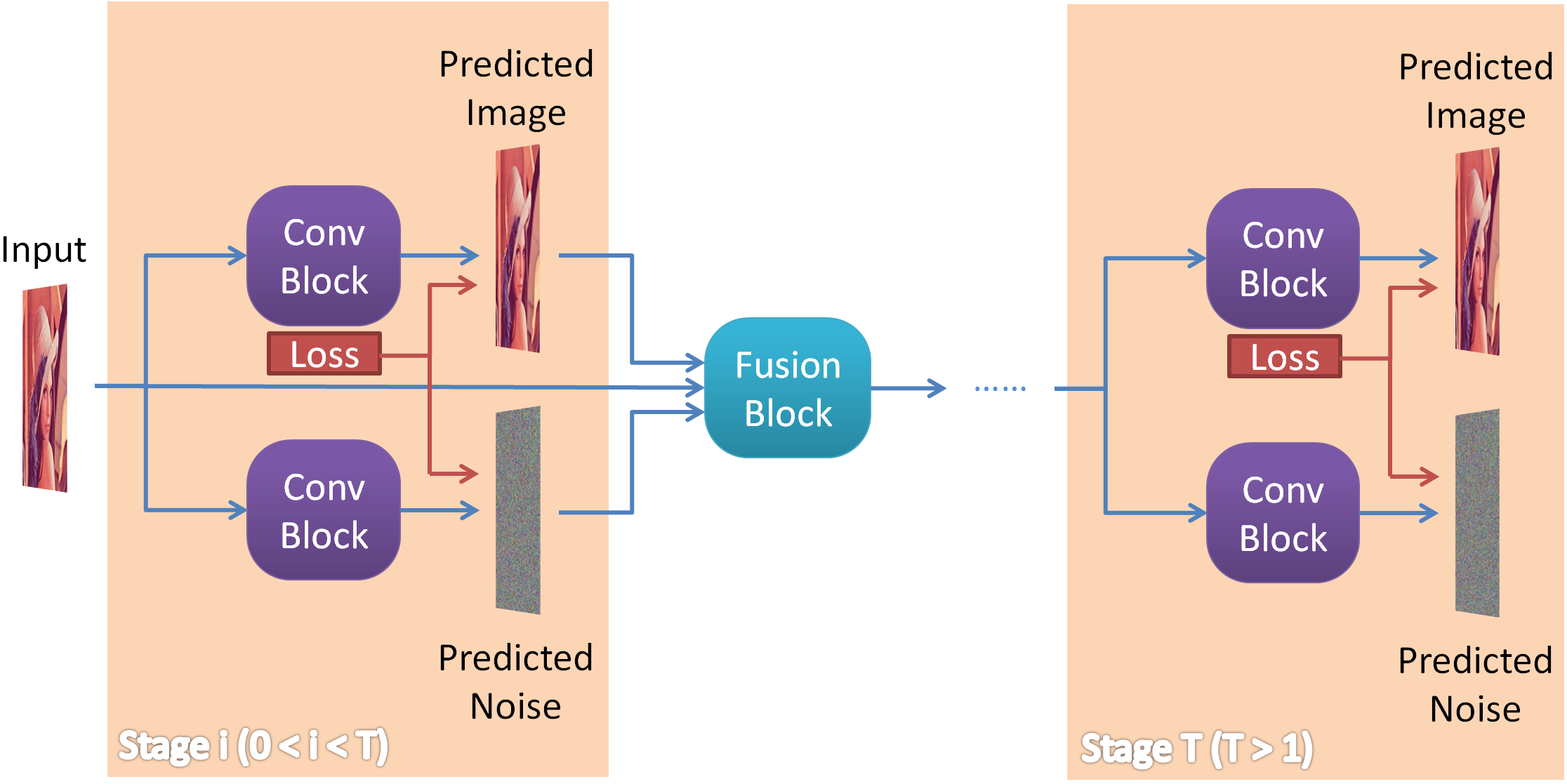}\\
  \caption{A high level concept of network architecture for the proposed NFCNN.}\label{high_level_network_architecture}
\end{figure}
\cref{high_level_network_architecture} illustrates the high level concept of network architecture for NFCNN. There are multiple stages contained in NFCNN to simultaneously output intermediate predicted noise and intermediate predicted clean images, and there is a fusion block between every two stages. Each fusion block blends the data of predicted noise, predicted clean image and original input together and outputs a fused result to next stage.

Let $\textbf{S}_i$ denote the i-th stage for $0<i\leq T$ and $T>1$. There are two branches in $\textbf{S}_i  $  to predict the latent clean image and noise by convolutional blocks, respectively.
\begin{figure}[!t]
  \centering
  \includegraphics[width=4in]{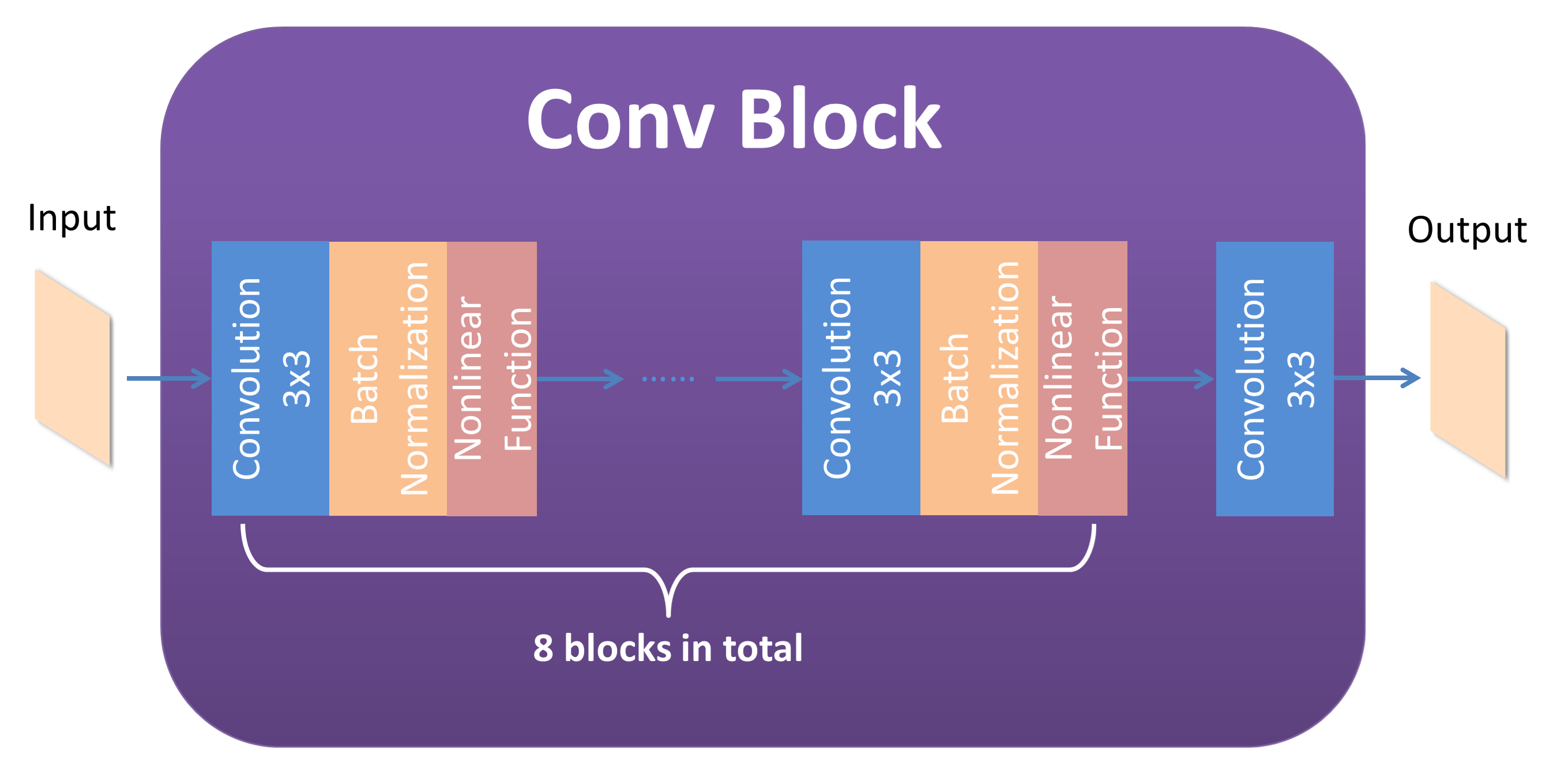}\\
  \caption{Detailed design for the convolutional block.}\label{conv_block}
\end{figure}
The detailed design for this convolutional block is shown in \cref{conv_block}. Nine convolutional layers with $3\times3$ kernel size are contained in the block. Following the design principle of DnCNN \cite{zhang2017beyond}, we set the batch normalization layer behind the convolutional layer, and the nonlinear activation function layer is next to the batch normalization layer. The last convolutional layer  directly generates the predict result. The number of channels for each convolutional layer is set to be 32 at first, then gradually reaches to 256 by doubling. It gradually returns to 32 at the last convolutional layer.

As demonstrated in \cref{high_level_network_architecture}, our proposed NFCNN has a phased network architecture, which may lead to a deep CNN model. Since vanishing gradient and exploding gradient problems might arise during the training process of a deep neural network, we employ a stage-wise supervised training policy to avoid these issues. The loss function is going to be calculated once the intermediate predicted results are generated from one stage. The loss of each stage will be summed up before the back propagation \cite{rumelhart1986learning}. Only the predicted clean image from the last stage of NFCNN is regarded as the final predicted result, because our NFCNN generalizes a predicted clean image better than a predicted residual image.

\subsection{Fusion Block}\label{subsection_fusion_block}
As displayed in \cref{high_level_network_architecture}, there is a fusion block between every two stages, except for the last stage, to achieve information fusion in our network. Though a few  methods use residual learning, i.e. let ANN learn the noise of image, they do not make enough use of noise. Information contained in noise is abundant and integrating it to network may help the model obtain a better result. To absorb the texture of image contained in noise, NFCNN takes the original noisy image, intermediate predicted clean image and intermediate predicted noise   as inputs of fusion block to generate a fused result of them. Thus, the fusion block can be regarded as an encoder and the next stage is treated as a decoder. The next stage directly takes the outputs from the fusion block as inputs to generate its outputs.
\begin{figure}[!t]
  \centering
  \includegraphics[width=4.5in]{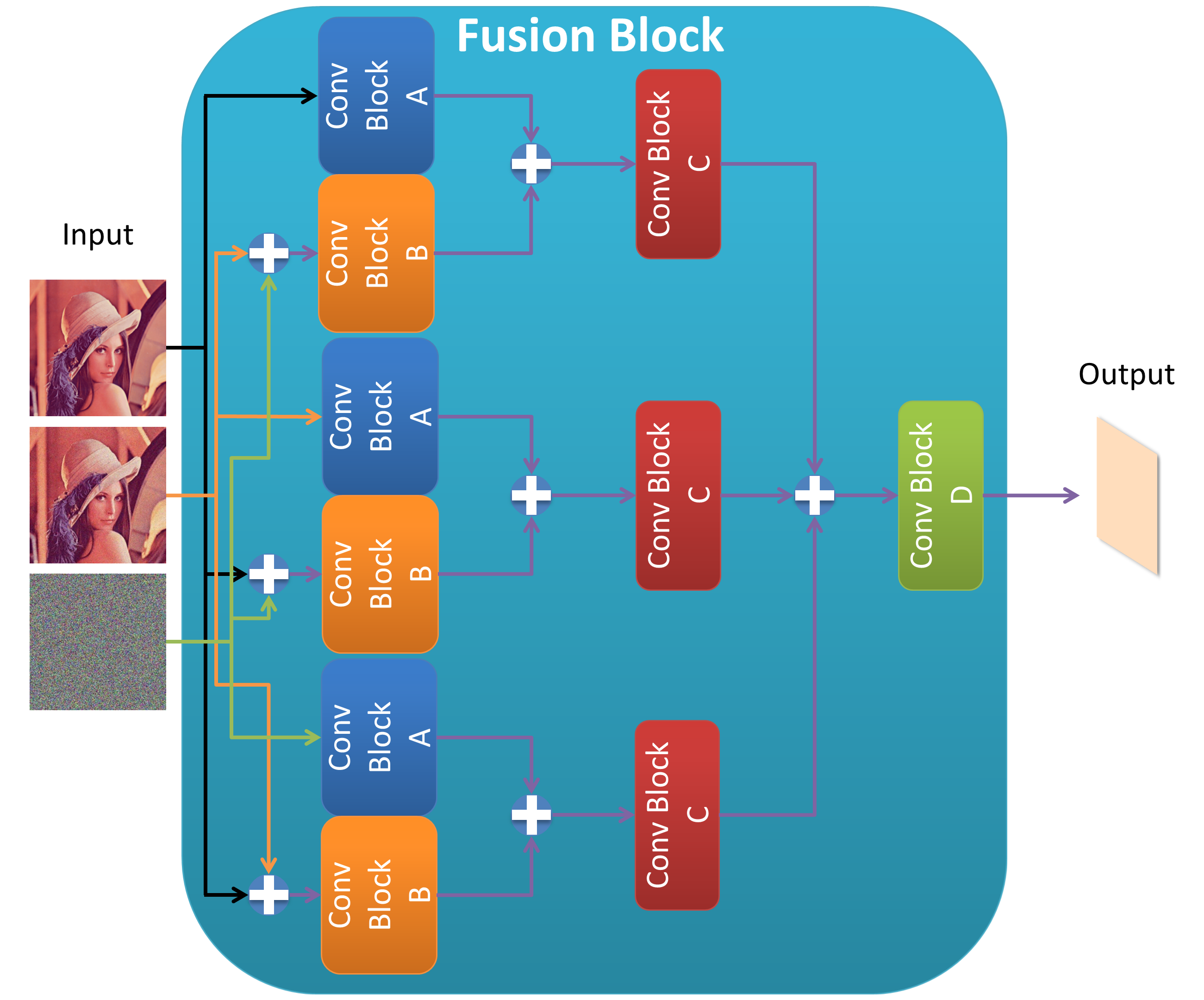}\\
  \caption{Architecture of fusion block. $\oplus$ denotes concatenating operation in channel axis.}\label{fusion_block}
\end{figure}
\begin{figure}[!t]
  \centering
  \includegraphics[width=4.5in]{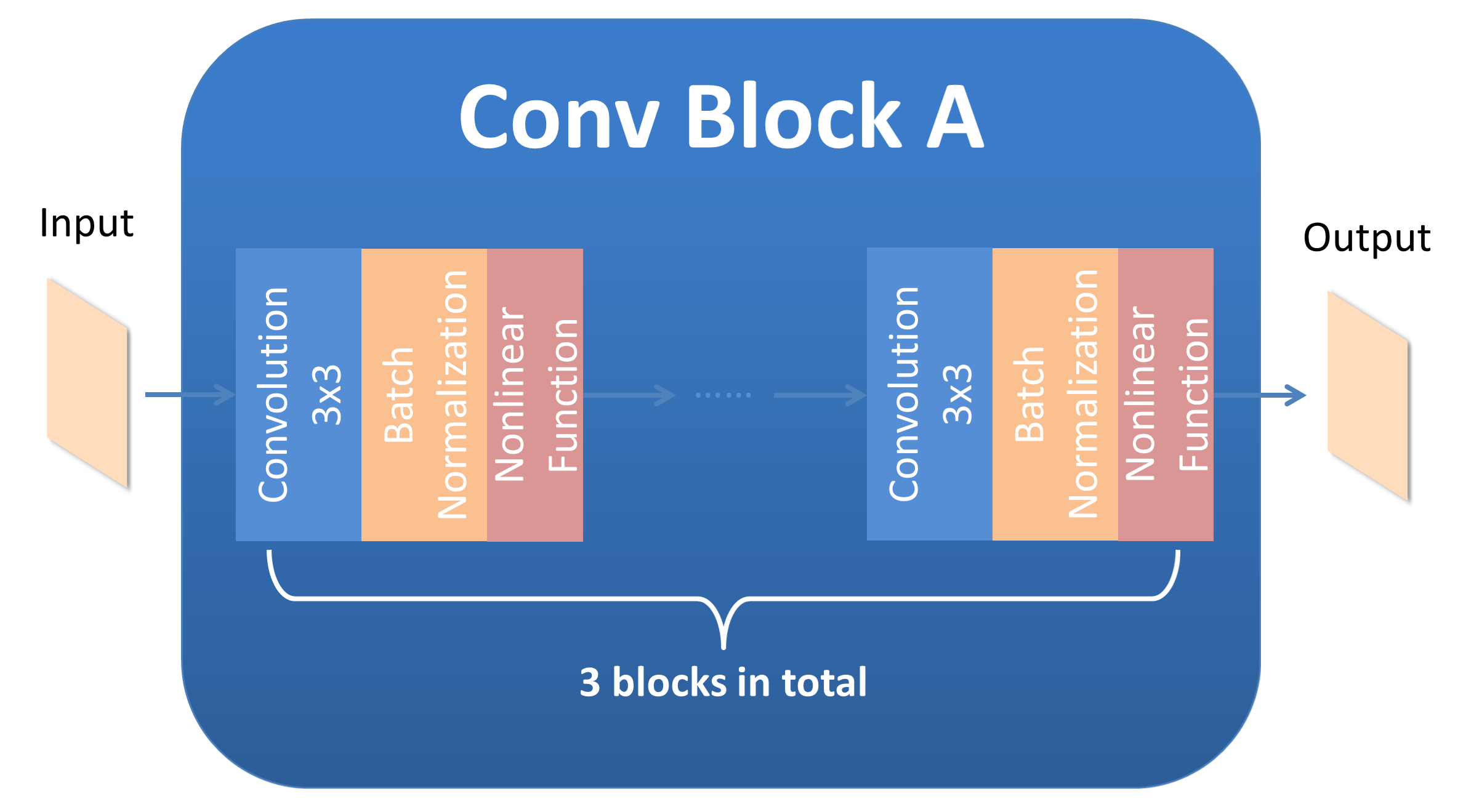}\\
  \caption{Detailed design for convolutional block A.}\label{conv_block_A}
\end{figure}
\begin{figure}[!t]
  \centering
  \includegraphics[width=4.5in]{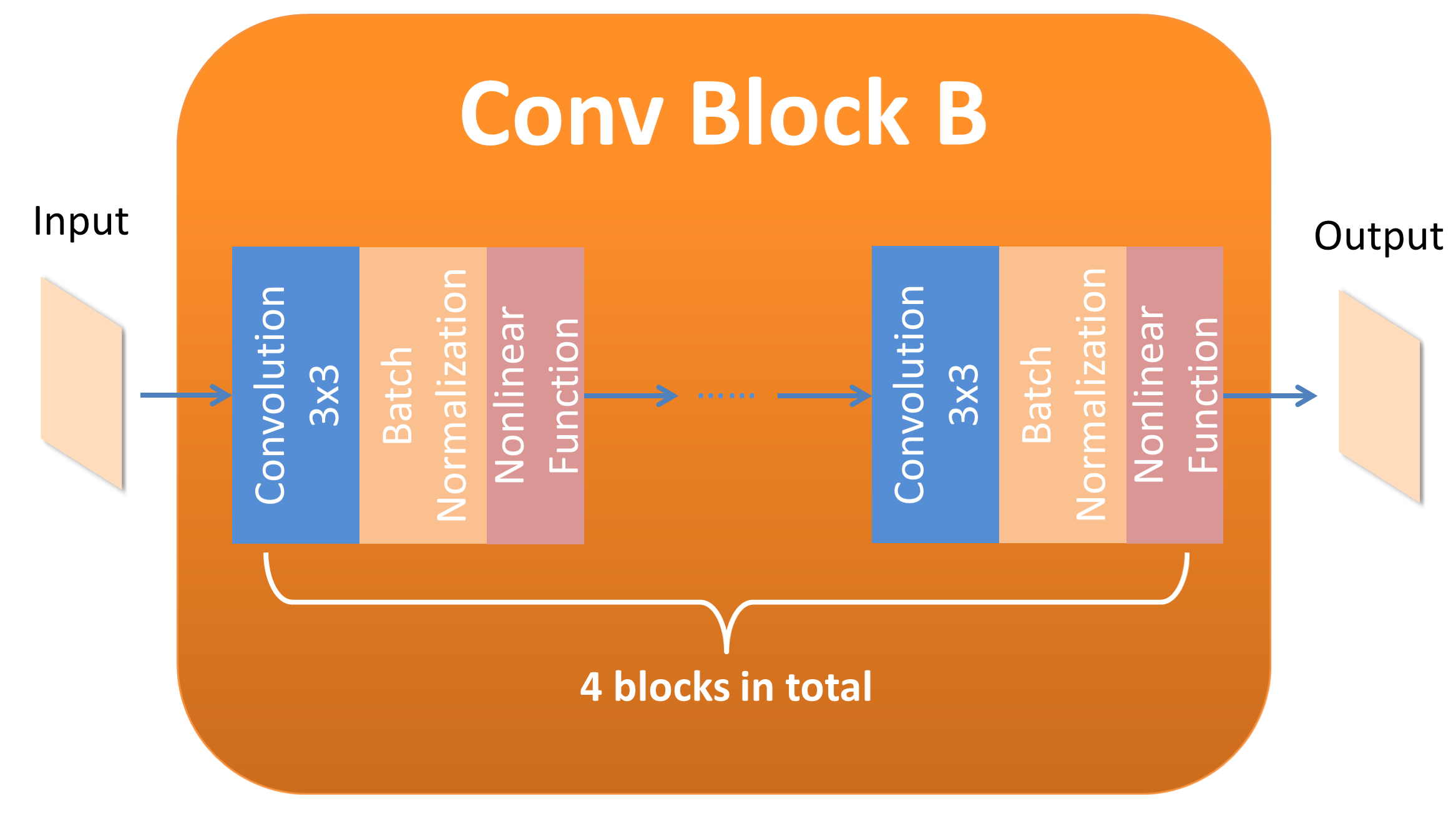}\\
  \caption{Detailed design for convolutional block B.}\label{conv_block_B}
\end{figure}
\begin{figure}[!t]
  \centering
  \includegraphics[width=4.5in]{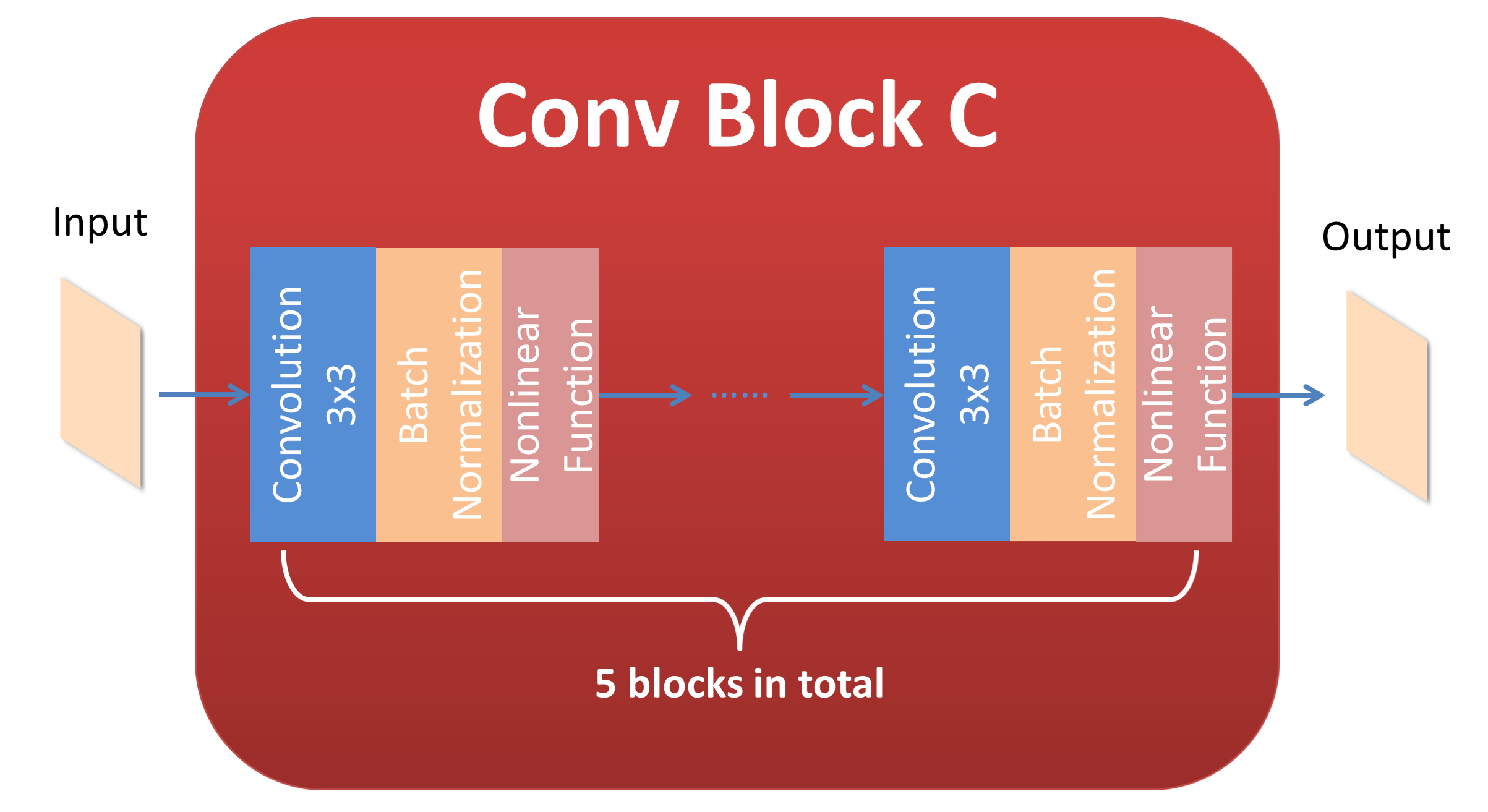}\\
  \caption{Detailed design for convolutional block C.}\label{conv_block_C}
\end{figure}
\begin{figure}[!t]
  \centering
  \includegraphics[width=4.5in]{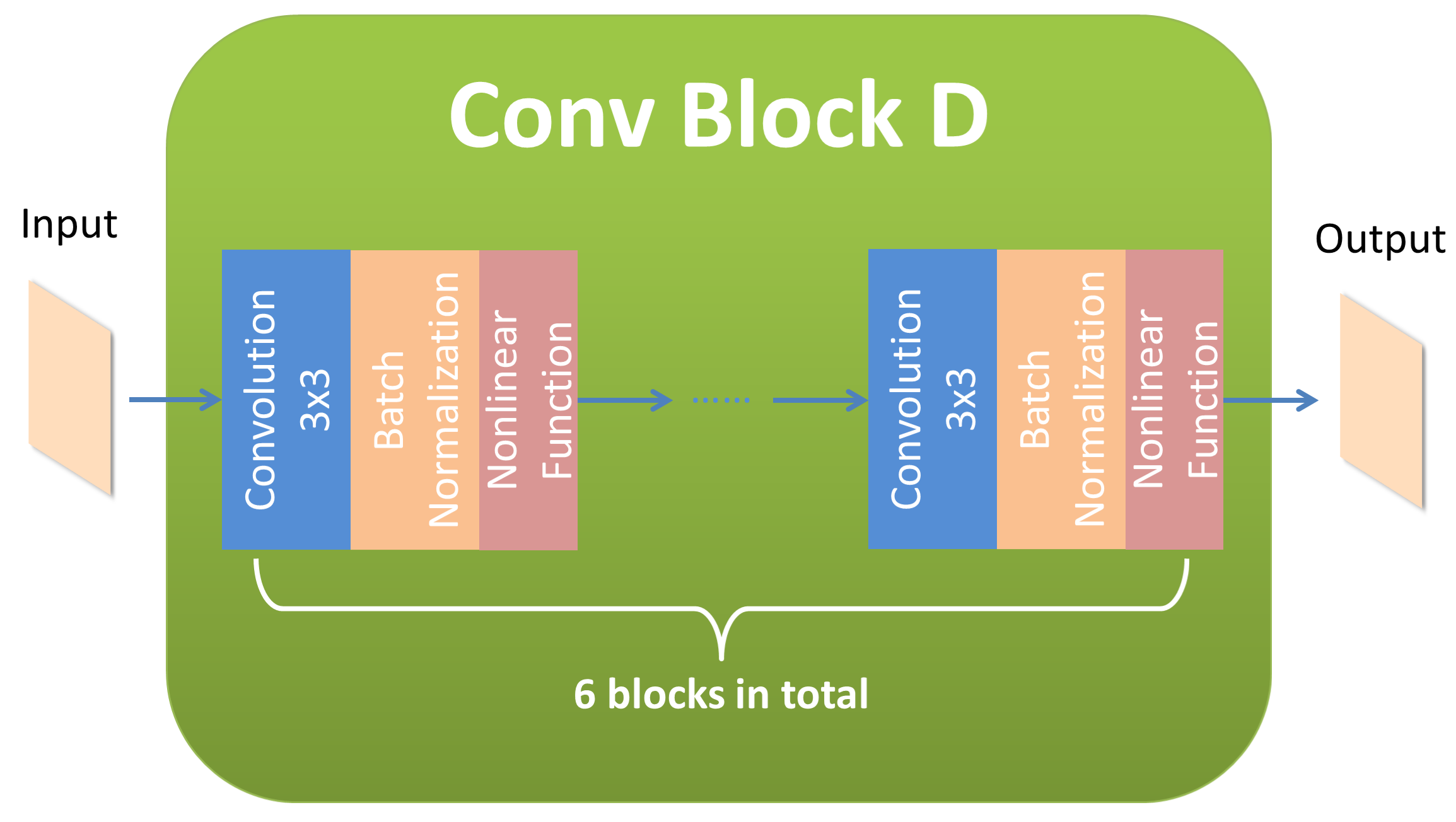}\\
  \caption{Detailed design for convolutional block D.}\label{conv_block_D}
\end{figure}

The architecture of fusion block is illustrated in \cref{fusion_block}, where $\oplus$ denotes concatenating operation in channel axis. In the fusion block of the  first layer, every two of inputs are combined without repetition as the input of convolutional block B, and the rest one is taken as the input of convolutional block A. The goal of first layer is to separately encode the predicted noise,  predicted clean image and original noisy image. Convolutional block A aims at  the encoding process of a single input  and convolutional block B is for the encoding process of a composed input. \cref{conv_block_A} and \cref{conv_block_B} show a detailed design that is similar to \cref{conv_block} for convolutional blocks A and B, respectively.

In the second layer, the encoded outputs from  the first layer are concatenated in channel as the input of convolutional block C, whose architecture is shown in \cref{conv_block_C}. Outputs from C are composed again before convolutional block D. The architecture of D is given by \cref{conv_block_D}. Note that the number of convolutional layers for convolutional blocks A, B, C and D is increasing from three to six gradually, so as to make the encoding process of noise become more abstract as the number of layers increases. By the fusion block, information contained in noise is able to be extracted and encoded in the outputs.

\subsection{Loss Function}
As mentioned before, our NFCNN has a deep structure due to the phased network architecture and fusion block. It is possible to encounter during  training some issues like vanishing gradient, exploding gradient, easily overfitting and being caught in a local minimum point, etc.. A possible way to avoid such situations is the stage-wise supervised training strategy.

The input of NFCNN is a noisy observation $\textbf{y}=\textbf{x}+n$, where $\textbf{x}$ is the clean image and $n$ denotes the noise.
The goal of our NFCNN is to learn a mapping $\mathcal{F(\textbf{y})=\hat{\textbf{x}}}$ to generate a latent clean image $\hat{\textbf{x}}$ from noisy observation $\textbf{y}$. Suppose $\textbf{F}_i$ to be the i-th fusion block, then the whole pipeline can be formulated by
\begin{equation}\label{NFCNN_0<i<_=_T}
\left\{
  \begin{array}{ll}
    \textbf{y}_1 = \textbf{y}, & \hbox{$i=0;$} \\
    (\textbf{C}_i, \textbf{N}_i) = \textbf{S}_i(\textbf{y}_i),~\textbf{y}_{i+1} = \textbf{F}_i(\textbf{y}_1, \textbf{C}_i, \textbf{N}_i), & \hbox{$0<i<T;$} \\
    (\textbf{C}_T, \textbf{N}_T) = \textbf{S}_T(\textbf{y}_T), & \hbox{$i=T,$}
  \end{array}
\right.
\end{equation}
where $\textbf{C}_i, \textbf{N}_i$ denote the intermediate predicted clean image and intermediate predicted noise, respectively.

The loss function of NFCNN is of the form
\begin{equation}\label{loss_function}
  \mathcal{L} = \mathcal{L}_C + \alpha\mathcal{L}_N,
\end{equation}
where $\alpha$ is a weight factor to balance  $\mathcal{L}_C$ and $\mathcal{L}_N$, the loss functions for intermediate predicted clean images and intermediate predicted noises, which are respectively given by
\begin{equation}\label{loss_function_C}
  \mathcal{L}_C(\mathbf{\Theta}) = \frac{1}{2K}\sum_{i=1}^{K}\nm{\mathcal{F}_C(\textbf{y}_i; \mathbf{\Theta}) - \mathbf{\hat{C}}_i}_2^2 + \beta\frac{1}{K}\sum_{i=1}^{K}\nm{\mathcal{F}_C(\textbf{y}_i; \mathbf{\Theta}) - \mathbf{\hat{C}}_i}_1,
\end{equation}
\begin{equation}\label{loss_function_N}
  \mathcal{L}_N(\mathbf{\Theta}) = \frac{1}{2K}\sum_{i=1}^{K}\nm{\mathcal{F}_N(\textbf{y}_i; \mathbf{\Theta}) - \mathbf{\hat{N}}_i}_2^2 + \beta\frac{1}{K}\sum_{i=1}^{K}\nm{\mathcal{F}_N(\textbf{y}_i; \mathbf{\Theta}) - \mathbf{\hat{N}}_i}_1.
\end{equation}
Here $\mathbf{\Theta}$ is the trainable parameters vector, $K$ the number of training samples, $\mathcal{F}_C$ the predicted clean image,  $\mathcal{F}_N$ the predicted noise, $\mathbf{\hat{C}}$ the ground-truth clean image, $\mathbf{\hat{N}}$ the ground-truth noise, and $\beta$ is a nonnegative factor. Note that the case of $\beta=0$ means that we purely use $L_2$ loss during training.

In the loss function \cref{loss_function}, we combine $L_2$ loss with $L_1$ loss to boost the generalization performance of NFCNN. Using $L_2$ loss during the training phase aims to match the labels of ground-truth clean image and noise in the sense of least squares, and $L_2$ loss is   related to PSNR metric to ensure the visual quality of our predicted results. As a contrast, employing $L_1$ loss in the loss function is to force the predicted result to approximate the label in the pixel level. However, since $L_1$ loss is related to SSIM metric, it is possible to reduce the convergence speed and performance of our model if $L_1$ loss occupies an excessive proportion. Hence, the factor $\beta$ is utilized  to control the influence brought from $L_1$ loss.

\subsection{Padding Mode}
Since the convolutional operation reduces the size of input, there are lots of modes for padding to keep the size same. The most common and popular mode is zero padding, which, however, might result in the artificial boundary effect or synthetic effect. Besides, zero padding is found to reduce the generalization performance of NFCNN, though it may be usable enough for higher computer vision tasks like image classification \cite{perronnin2010improving, krizhevsky2012imagenet, simonyan2014very}, pose estimation \cite{cao2018openpose, newell2016stacked, toshev2014deeppose} and object detection \cite{redmon2016you, he2017mask, lin2017focal}. As a result, we apply the replication padding mode in our method, whose core idea is to copy and expand the boundary values rather than just adding zero values.

\section{Experimental Results}\label{section4}
\subsection{Datasets}
To train our NFCNN with adequate data, we adopt some existing datasets to enrich our training data.
\begin{itemize}
\item \emph{BSDS500} \cite{amfm_pami2011}. BSDS500 dataset from Berkeley is contained in our training dataset. There are 500 color images that are originally designed for the contour detection task contained in this dataset. For an image denoising task, only the clean images are needed. Therefore, the ground-truth samples of this dataset are suitable for our task.
\item \emph{Waterloo Exploration Database} \cite{ma2017waterloo}. The dataset has 4,744 images with high quality from the Waterloo Exploration Database. This large-scale dataset is for testing the generalization performance of image quality assessment (IQA) models.
\item \emph{Flickr2K} \cite{lim2017enhanced}. The last part of our training data comes from Flickr2K dataset. It has 2,650 2K images for training of super-resolution task. However, 2K resolution for the image denoising task is too large to train our network. Thus, we crop them with a fixed patch size.
\end{itemize}
Note that the test samples are not contained in the training dataset. The whole training dataset contains 7,794 samples, and we adopt the validation set of BSDS500 as our validation dataset.

\subsection{Data Augmentation}
To avoid overfitting as far as possible, some data augmentation methods are employed in the training of NFCNN for both gray images and color scale images.
\begin{itemize}
\item \emph{Randomly Cropping}. Input for NFCNN is randomly cropped by a fixed patch size, e.g. $180\times180$, to prevent overfitting happening, though there are no constraints on its size. This operation also keeps a good balance between the GPU memory and training efficiency.
\item \emph{Flipping}. Input will be flipped in terms of three modes, i.e. horizontal flipping, vertical flipping and the combination mode of the former two.  By this data augmentation method, NFCNN can learn different patterns of image and noise.
\item \emph{Image Blurring}. The image blurring augmentation method is applied before adding the noise to the original clean image, which aims to simulate the issue of shooting jitter.
\end{itemize}

\subsection{Effectiveness of Fusion Block}
To show the effectiveness of fusion block, experiments are conducted by deleting the  fusion block between two neighbouring stages. NFCNNs with 2, 3 and 4 stages are regarded as experimental target models. According to \cref{subsection_fusion_block}, information contained in predicted clean image and predicted noise can be exchanged through the fusion block. Noise level $\delta$ is set to 15, 25, 50 and 75 during the experiments. Kodak24 dataset is employed as the test dataset to verify the effectiveness of fusion block.
\begin{table*}[!t]
  \caption{Results of different noise levels on Kodak24.}
  \label{effectiveness_of_fusion_block}
  \centering
  \scalebox{0.7}{
  \begin{tabular}{|c|c|c|c|c|c|c|c|c|c|c|c|}
  \hline \multicolumn{1}{|c|}{} & \multicolumn{2}{c|}{$\delta=15$} & \multicolumn{2}{c|}{$\delta=25$} & \multicolumn{2}{c|}{$\delta=50$} & \multicolumn{2}{c|}{$\delta=75$} \\
  \hline {Number of Stages} & {NFCNN($\ast$)} & {NFCNN} & {NFCNN($\ast$)} & {NFCNN} & {NFCNN($\ast$)} & {NFCNN} & {NFCNN($\ast$)} & {NFCNN} \\
  \hline
  {2} & {34.54} & {\textbf{34.72}} & {31.95} & {\textbf{32.15}} & {28.23} & {\textbf{28.91}} & {26.55} & {\textbf{27.29}}\\
  \hline
  {3} & {32.65} & {\textbf{33.17}} & {30.46} & {\textbf{31.06}} & {26.57} & {\textbf{27.18}} & {25.86} & {\textbf{26.88}}\\
  \hline
  {4} & {31.39} & {\textbf{32.48}} & {30.13} & {\textbf{30.71}} & {25.01} & {\textbf{25.54}} & {24.34} & {\textbf{24.63}}\\
  \hline
  \end{tabular}
  }
\end{table*}

\begin{figure*}[!t]
  \centering
  \subfloat{\includegraphics[width= 4in]{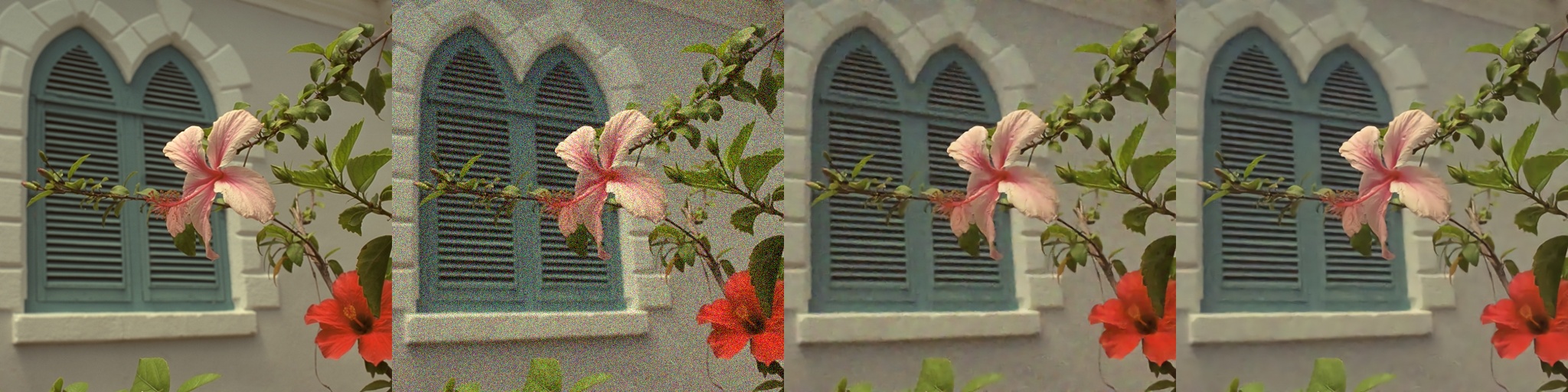}}\\
  \subfloat{\includegraphics[width= 4in]{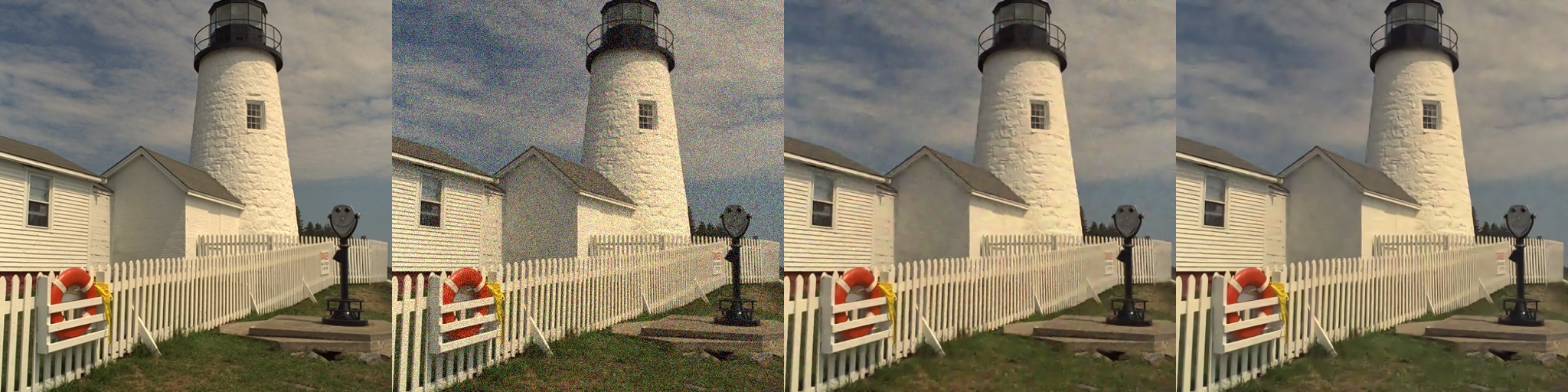}}\\
  \subfloat{\includegraphics[width= 4in]{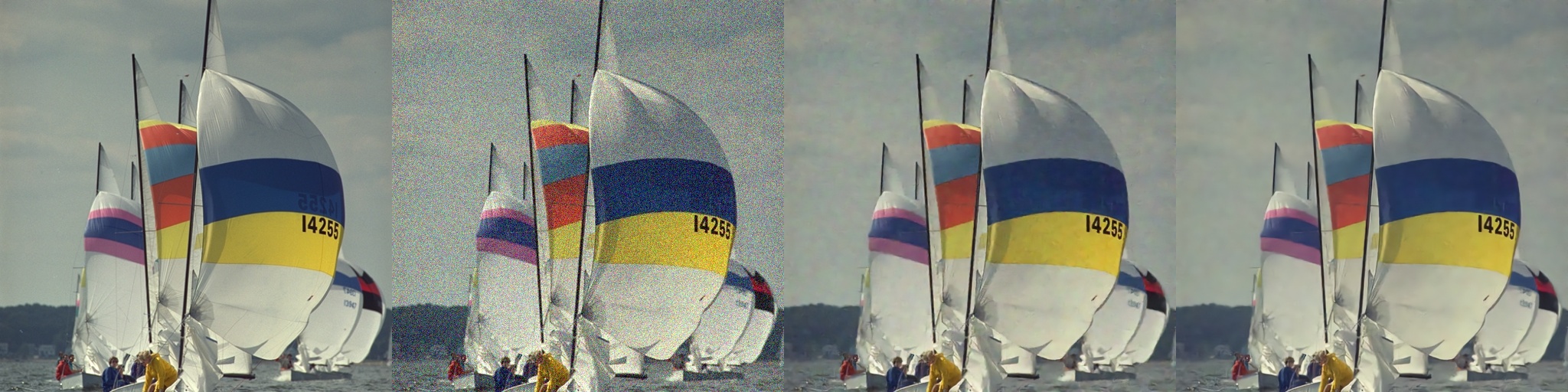}}\\
  \subfloat{\includegraphics[width= 4in]{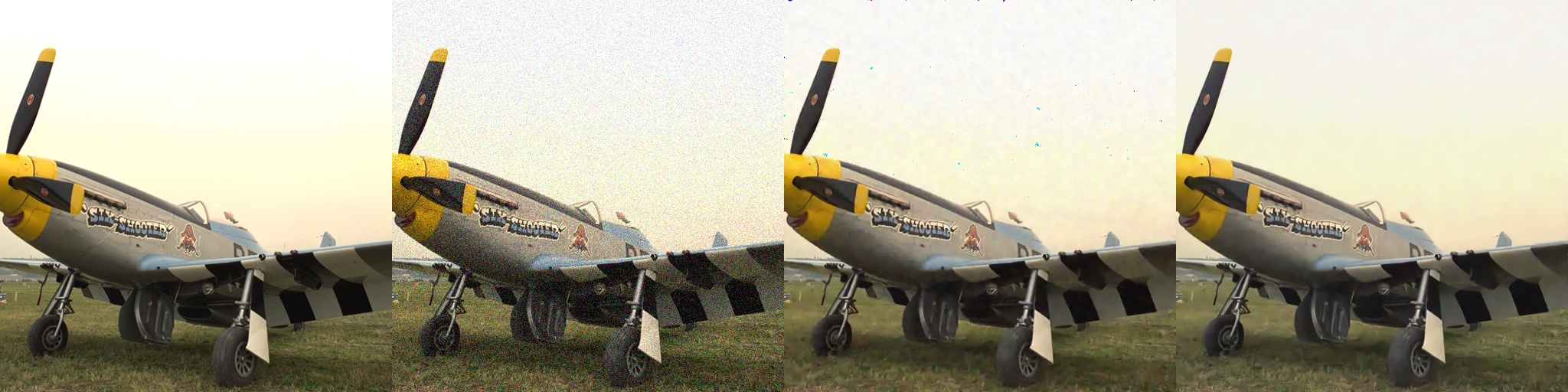}} \\
  \caption{Visualization results about the effectiveness of fusion block under noise level $\delta=25$, where the first, second, third and last columns are original images, noisy images, results of NFCNN($\ast$) and results of NFCNN, respectively.}\label{comparison_of_effectiveness_of_fusion_block}
\end{figure*}

Experimental results are displayed by \cref{effectiveness_of_fusion_block}, where NFCNN($\ast$) denotes the NFCNN model without the fusion block, with metric PSNR(dB). We can see that NFCNN outperforms NFCNN($\ast$), i.e. NFCNN with the fusion block behaves better than the one   without the fusion block. A few visualization results in \cref{comparison_of_effectiveness_of_fusion_block} also show that denoised results are refined by the fusion block. This means that the fusion block is helpful for the exchange of information between the predicted clean image and predicted noise and has the ability to decrease artificial effect. We also see that NFCNN has better performance when the number of stages of NFCNN is set as 2. Based on this observation, we set it as 2 in all the experiments for NFCNN.

\subsection{Experimental Setting and Network Training}
During training, we set $\alpha=1.0$ in \cref{loss_function} and  $\beta=0.01$ in \cref{loss_function_C} and \cref{loss_function_N} to control the influence from $L_1$ loss. LeakyReLU\cite{maas2013rectifier} with slope 0.25 is utilized in NFCNN as the nonlinear activation function. According to the receptive field, the number of stages in NFCNN should be set as an appropriate value to keep good balance between the size of receptive field and the patch size $180\times180$. Experiments of different numbers of stages have been conducted and the results are displayed in \cref{effectiveness_of_fusion_block}.

\begin{figure}[!t]
  \centering
  \subfloat[Original image]{\includegraphics[width= 1in]{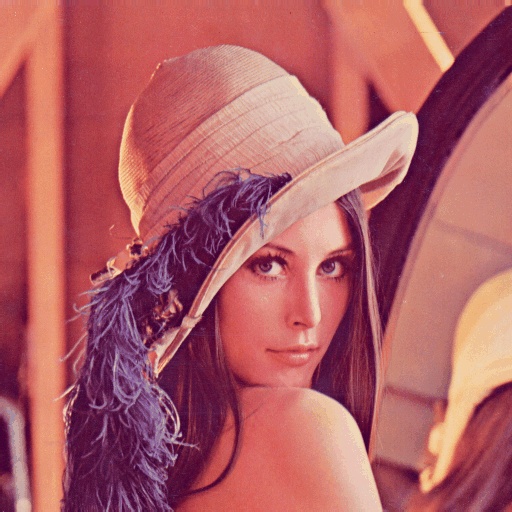}}\quad
  \subfloat[Noisy image with cropping operation]{\includegraphics[width= 1in]{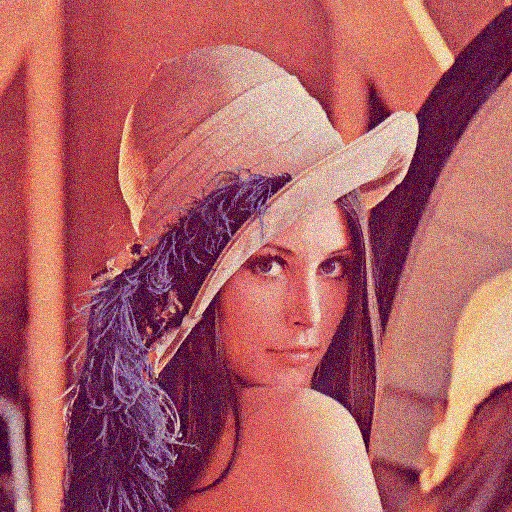}}\quad
  \subfloat[Noisy image without cropping operation]{\includegraphics[width= 1in]{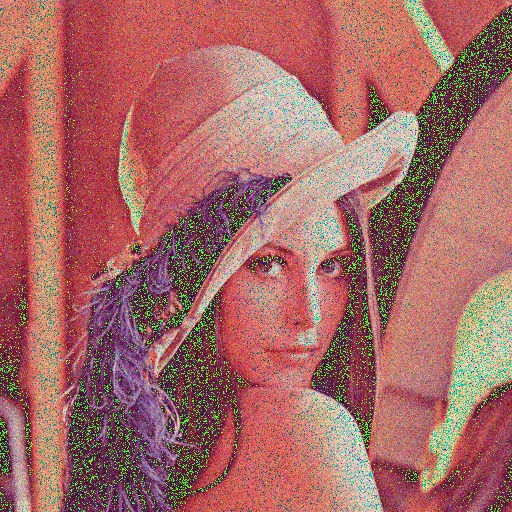}} \\
  \caption{Comparison of synthetic data with and without cropping operation when noise level $\delta=25$.}\label{comparison_of_cropping_and_noncropping}
\end{figure}

To simulate the noisy input, we apply Additive White Gaussian Noise (AWGN) with standard deviation $\delta$ to generate synthetic data for training. When using the normal distribution with cropping operation, i.e. cropping value of image to keep it in range $[0, 255]$, synthetic data can be more realistic. \cref{comparison_of_cropping_and_noncropping} shows the difference between the two patterns on Lenna image with or without cropping operation.

To test our NFCNN under different noise levels, we set the noise level parameter $\delta= 15, 25, 50, 75$ for different denoising tasks. Batch size is set as 6 to keep the balance between the training speed and GPU memory.

To make our NFCNN have a faster convergence speed, we employ Adam optimizer \cite{kingma2014adam} with learning rate $1e^{-3}$ and default setting of other parameters. Learning rate will be divided by 10 after 300,000 steps of training. The kernel size of all convolutional layer is set as $3\times3$ in the case of patch size $180\times180$, and it can be modified if the patch size becomes bigger. Dropout1d \cite{srivastava2014dropout} is contained in NFCNN to boost performance in the pixel level, and dropout2d \cite{tompson2015efficient} aims to decrease the relevancy in the channel level. Note that these two kinds of dropout methods are able to alleviate overfitting.

\subsection{Comparison with Different Models}
In this subsection, we compare our model with some state-of-the-art methods such as BM3D \cite{dabov2007image}, WNNM \cite{gu2014weighted}, MLP \cite{burger2012image}, TNRD \cite{chen2016trainable}, DnCNN \cite{zhang2017beyond}, and FFDNet \cite{zhang2018ffdnet}. Experiment results are listed in Tables \ref{BSD68_exp}-\ref{CMcMaster_exp} to demonstrate the competitive performance of our NFCNN. Since some methods are not open source, we can only borrow results from their corresponding original papers. Note that we do not compare NFCNN with CBDNet \cite{guo2019toward}  which is applicable to real-world denosing tasks.

For the BSD68 test dataset, all the algorithms are tested by adding noise levels $\delta=15, 25, 50$ and 75 for gray scale images. Results based on PSNR(dB) are displayed in \cref{BSD68_exp}. We can see that our proposed NFCNN with 2 stages behaves well in all cases. NFCNN surpasses FFDNet by 0.16dB when noise level $\delta=15$. NFCNN outperforms DnCNN when  $\delta=$ 15, 25 and 75, while they have similar generalization performance when $\delta=50$.
The reason for the  similar performance when $\delta=50$ may be that DnCNN has reached during its training a better minimum point in this case than that in other cases.

CBSD68 is a color version of BSD68, and the corresponding results are shown in \cref{CBSD68_exp}. When noise level $\delta= 15$ and 25, NFCNN exceeds DnCNN by 0.01$\thicksim$0.02dB at $\delta=$ 15 and 25, and DnCNN outperforms FFDNet by 0.02dB. However, NFCNN is defeated by FFDNet when $\delta=50$.

Set12 is another well known dataset for image denoising. According to \cref{Set12_exp}, NFCNN surpasses FFDNet by 0.03$\thicksim$0.04dB when $\delta=$ 50 and 75, and behaves better than DnCNN and FFDNet when $\delta=15$. When $\delta=25$, DnCNN and FFDNet have similar denoising performance and surpass NFCNN by 0.01dB.

\cref{CKodak24_exp} gives the average PSNR(dB) of different methods on Kodak24 dataset. We can see that in the cases of $\delta=$ 15, 25 and 75, NFCNN still has the best performance in terms of PSNR, but FFDNet has a better result when $\delta=50$.

\cref{CMcMaster_exp} shows the results on McMaster dataset. Thanks to the use of fusion block, NFCNN has the best performance. Note NFCNN reaches the same performance as that of FFDNet when $\delta=50$.

\begin{figure}[!t]
  \centering
  \subfloat{\includegraphics[width= 2in]{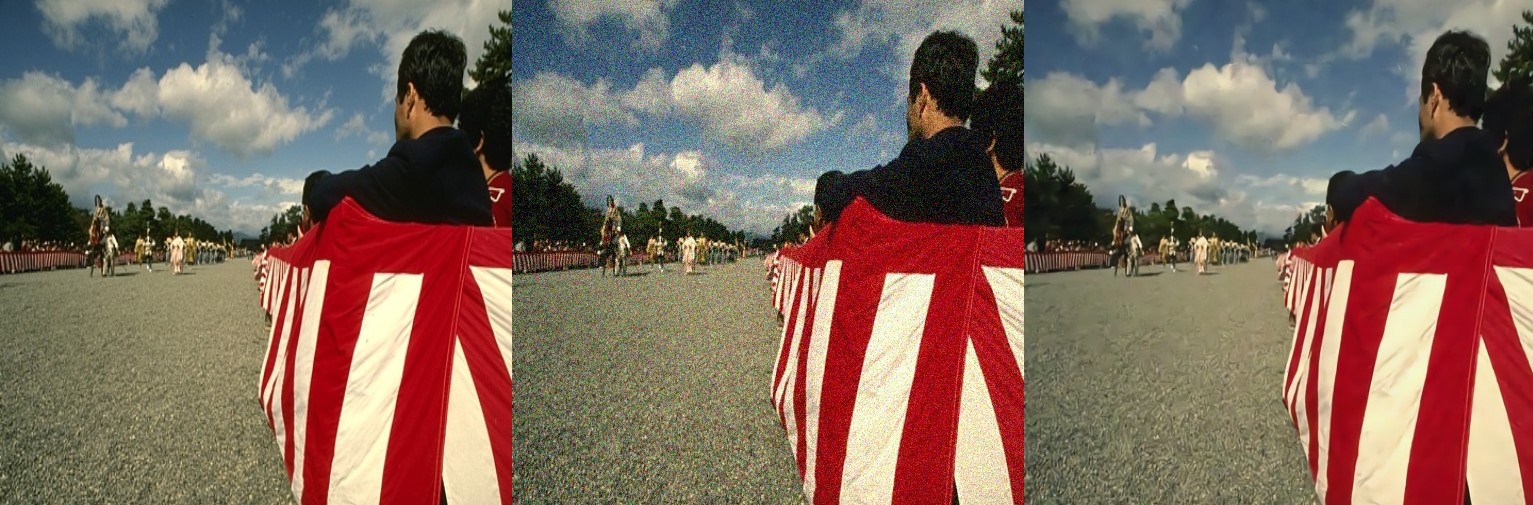}}\quad
  \subfloat{\includegraphics[width= 2in]{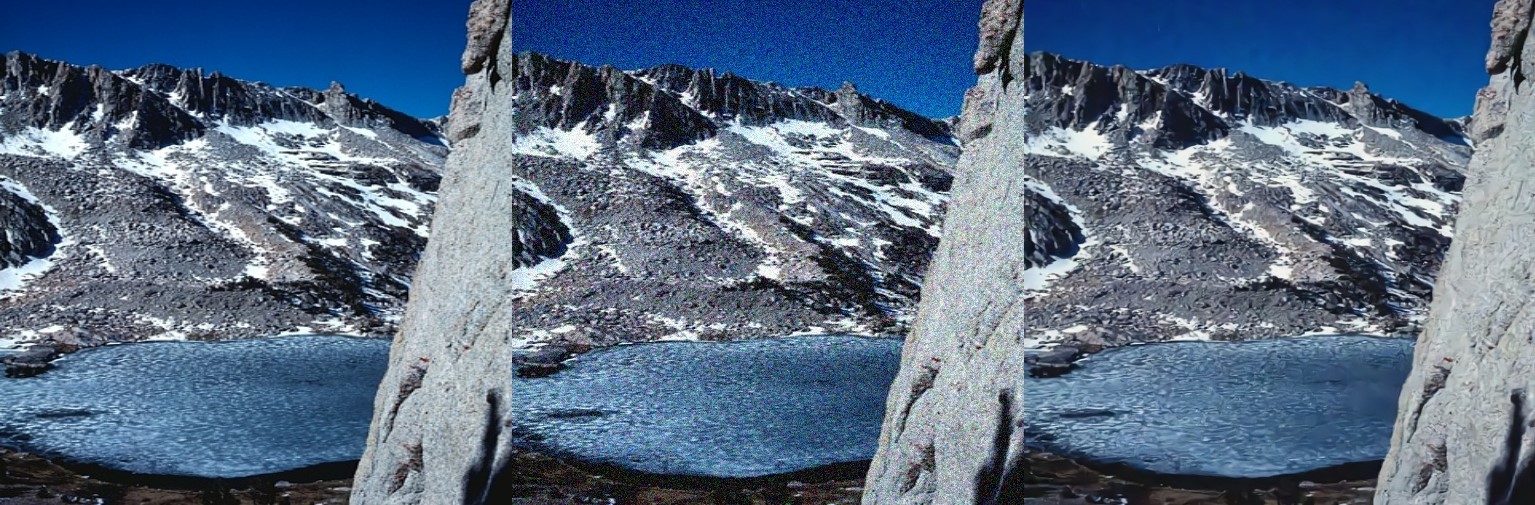}}\quad
  \subfloat{\includegraphics[width= 2in]{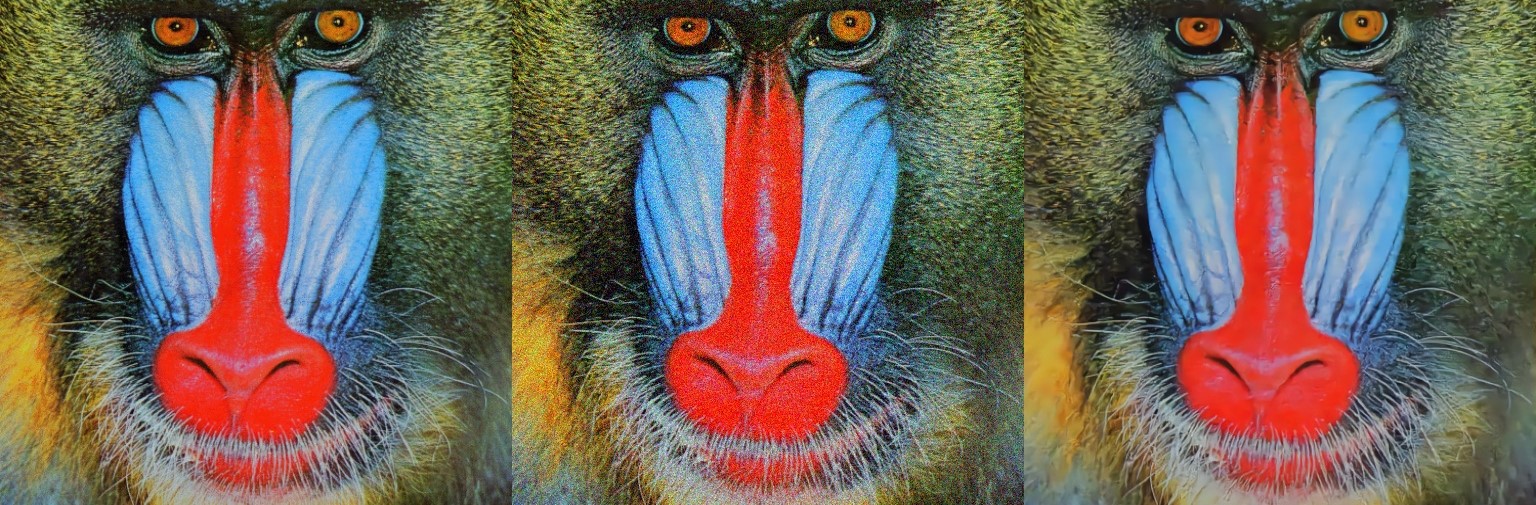}}\quad
  \subfloat{\includegraphics[width= 2in]{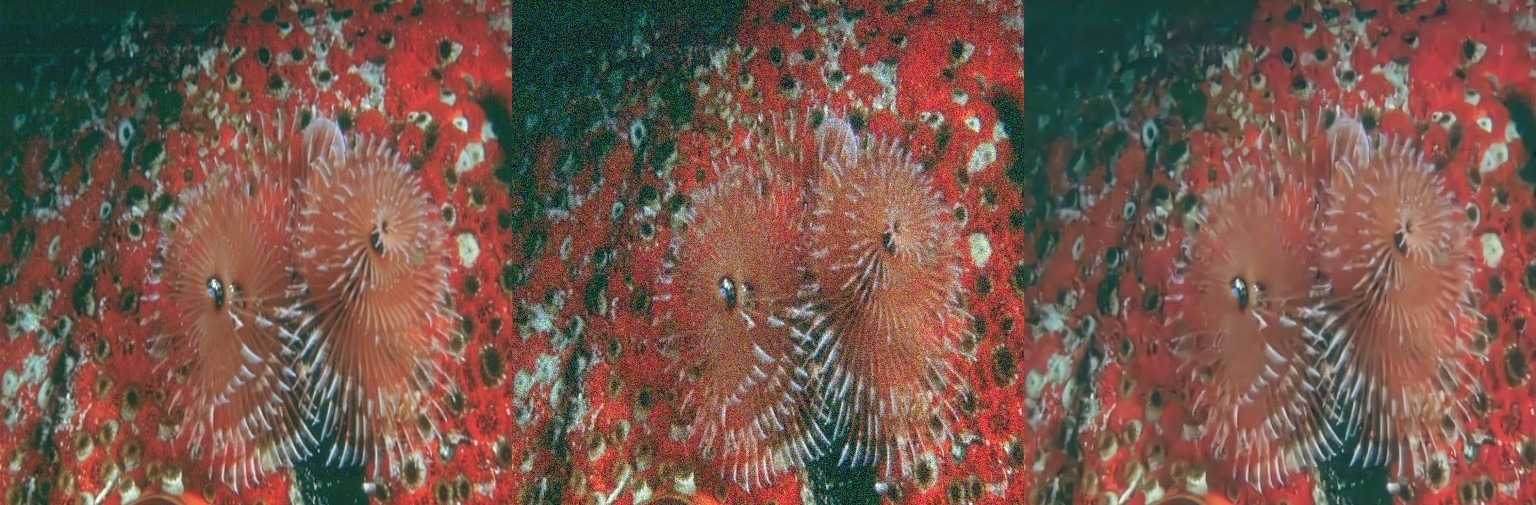}}\quad
  \subfloat{\includegraphics[width= 2in]{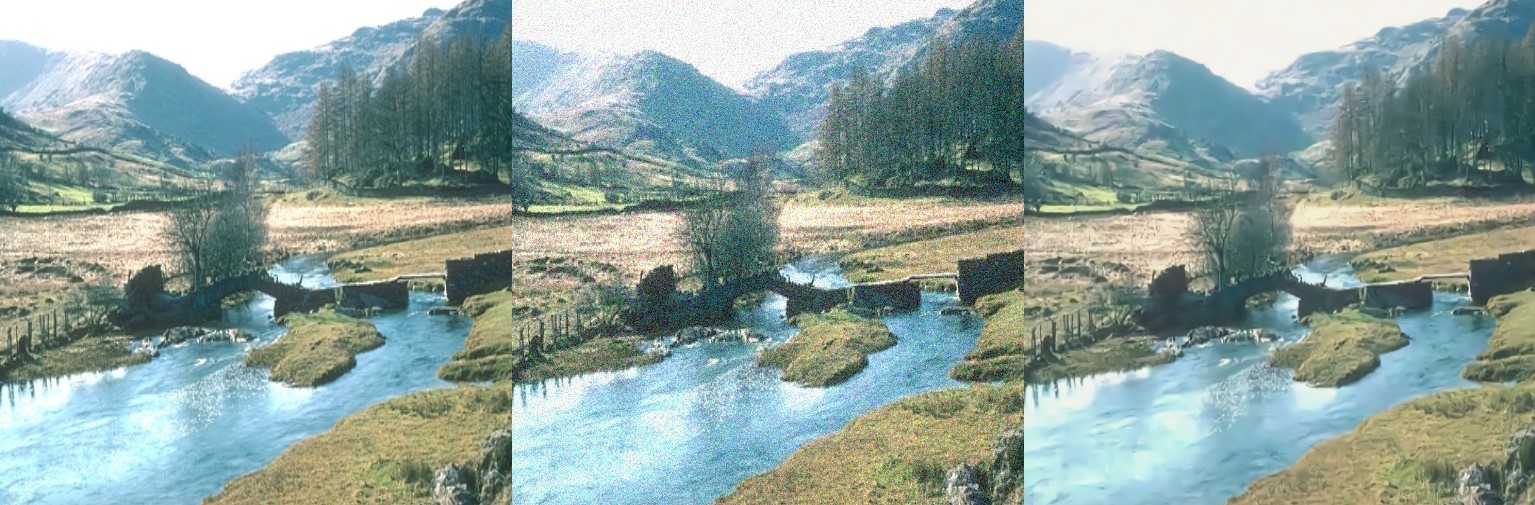}}\quad
  \subfloat{\includegraphics[width= 2in]{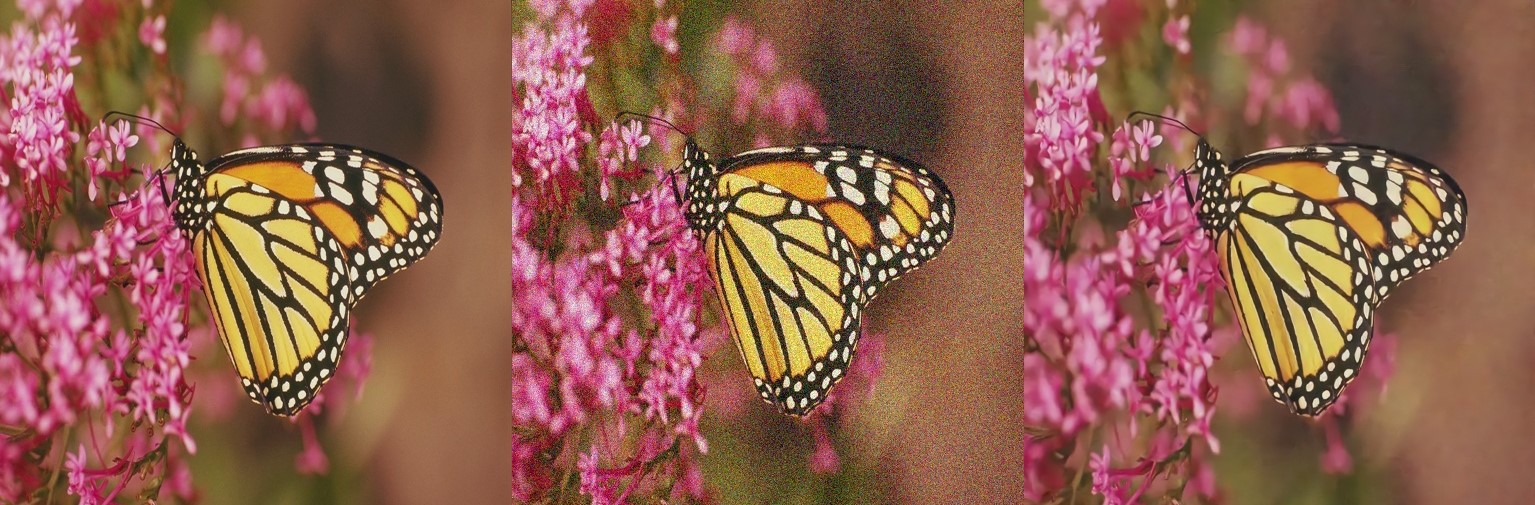}}\quad
  \subfloat{\includegraphics[width= 2in]{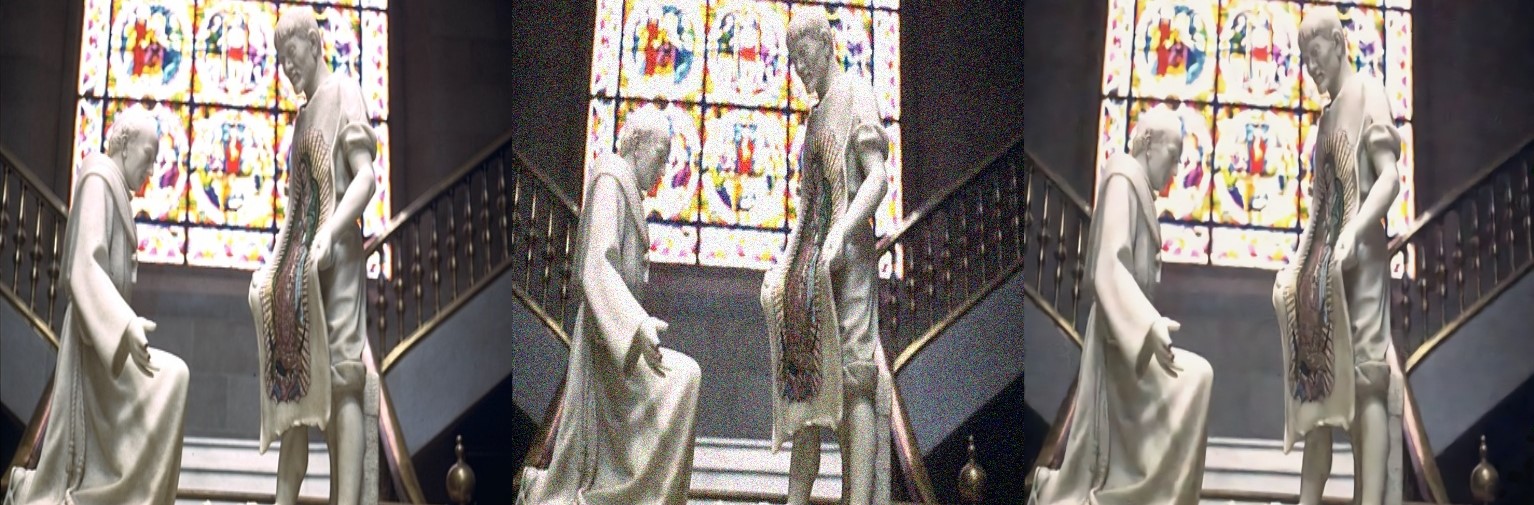}}\quad
  \subfloat{\includegraphics[width= 2in]{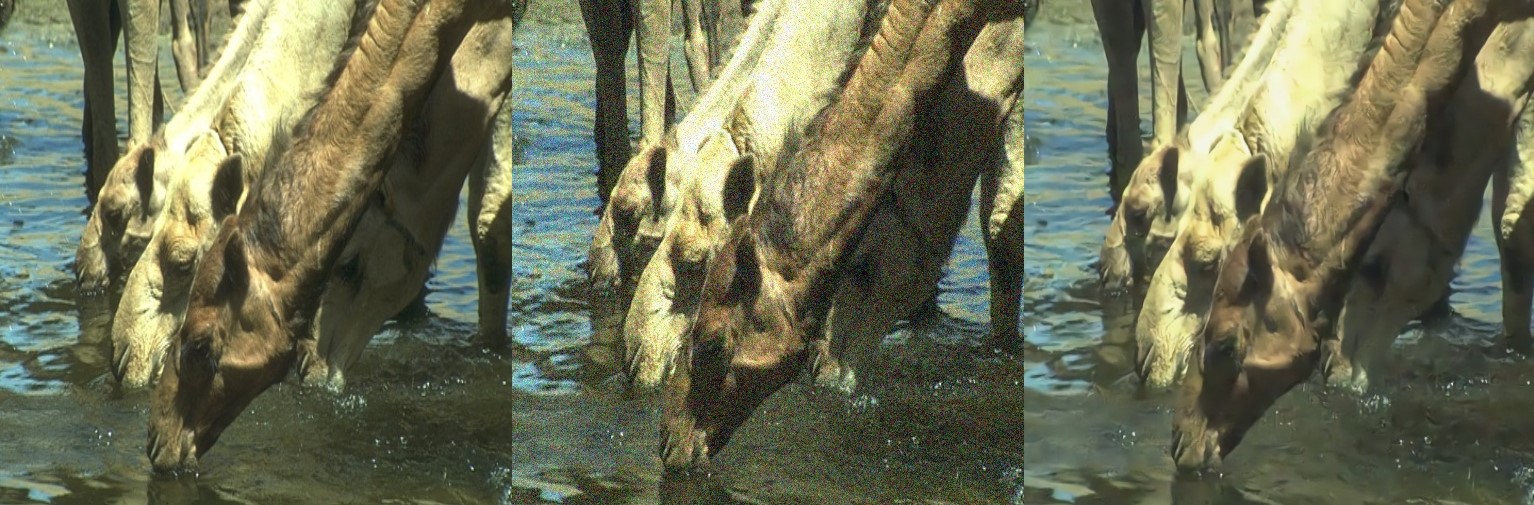}}\quad
  \subfloat{\includegraphics[width= 2in]{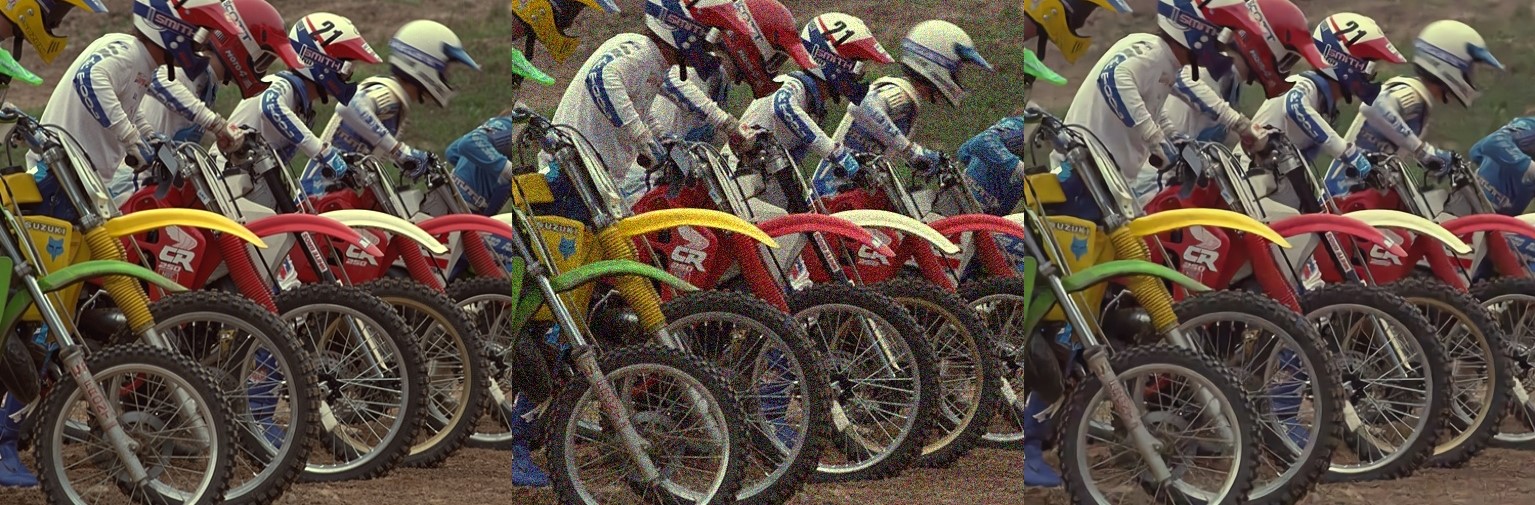}}\quad
  \subfloat{\includegraphics[width= 2in]{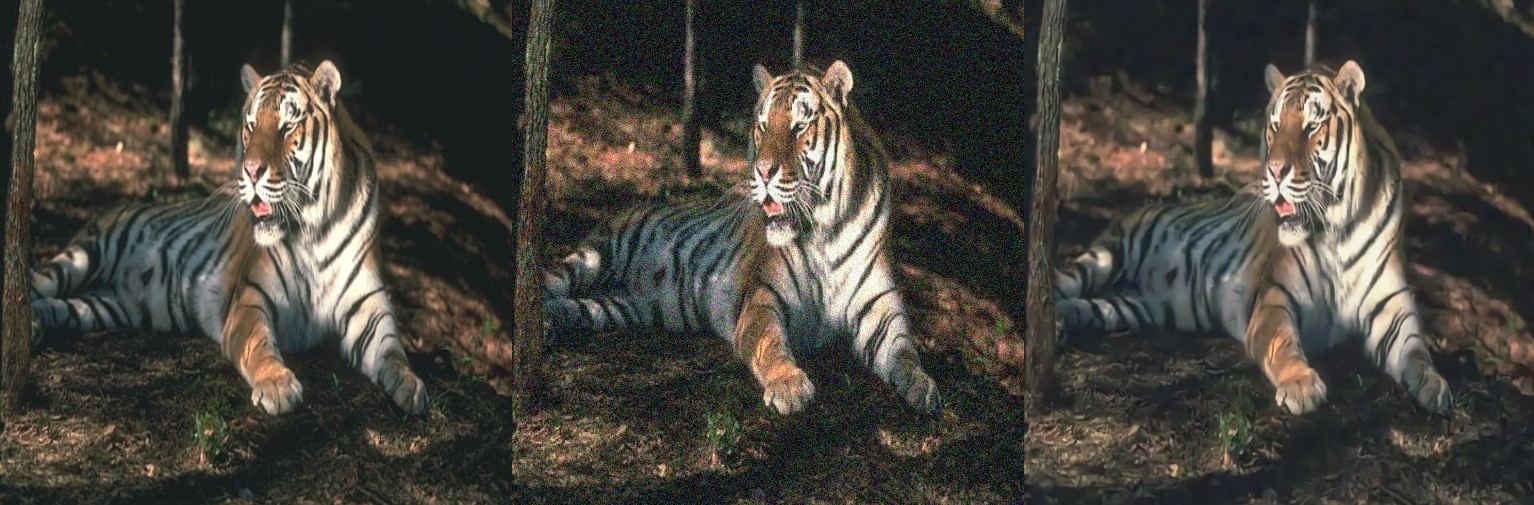}}\quad
  \subfloat{\includegraphics[width= 2in]{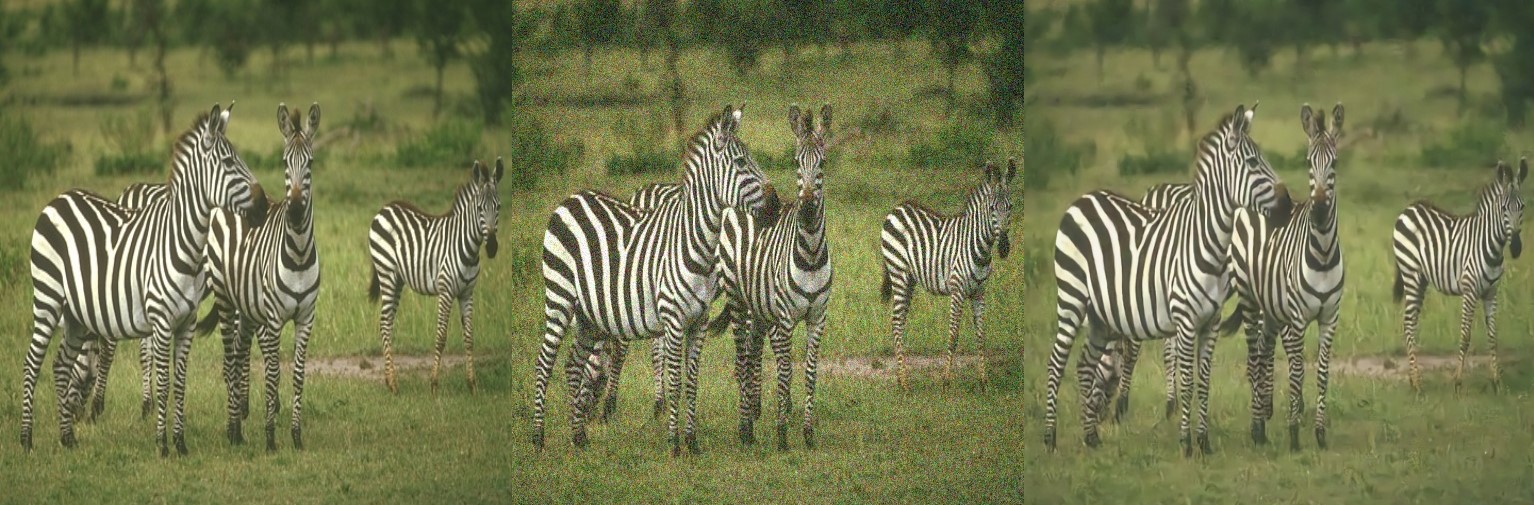}}\quad
  \subfloat{\includegraphics[width= 2in]{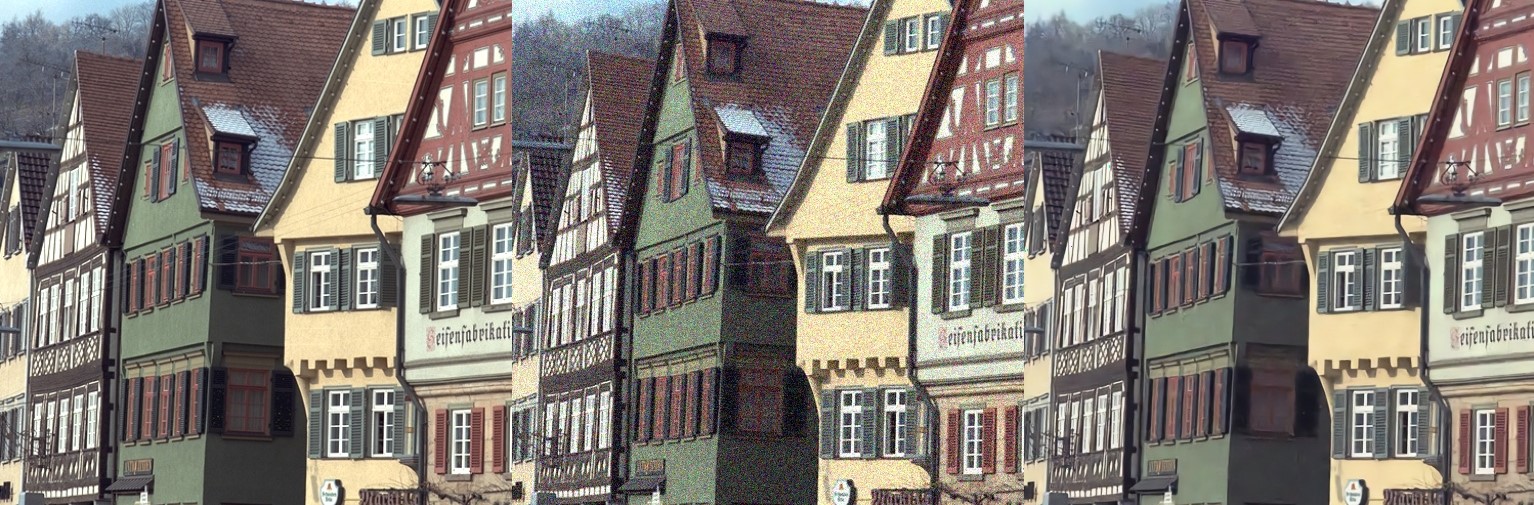}}\quad
  \subfloat{\includegraphics[width= 2in]{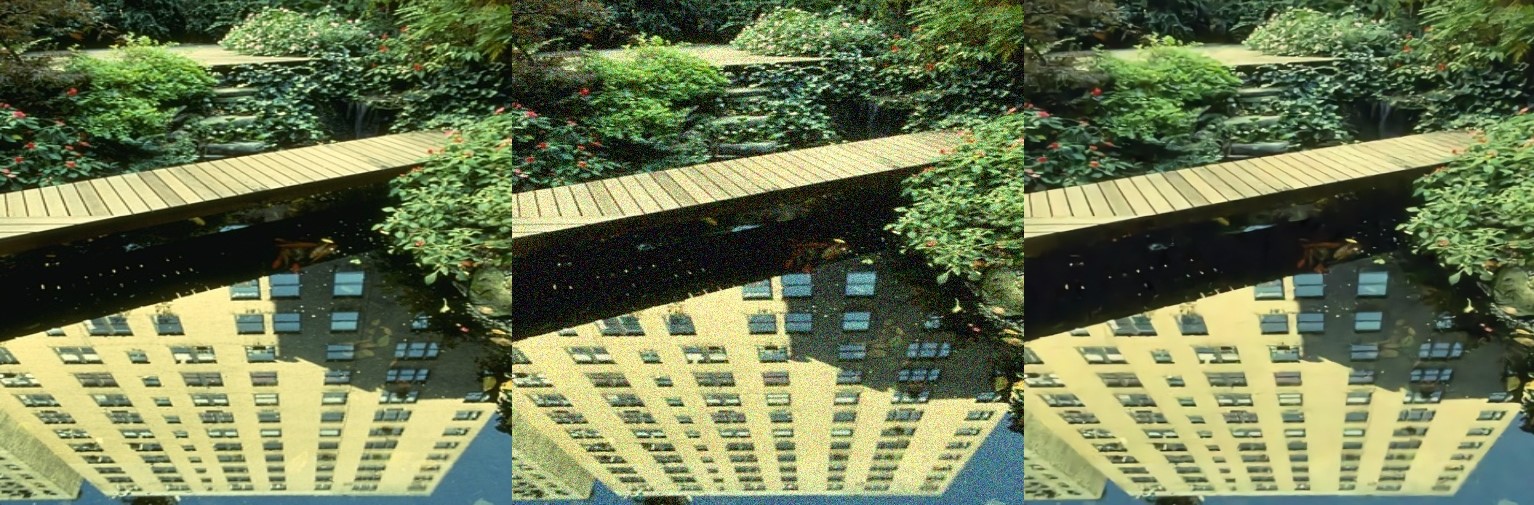}}\\
  \caption{Results of NFCNN with noise level $\delta=25$. The first, second and last columns are the  original images,   noisy images and denoised results, respectively.}\label{delta_25_results}
\end{figure}

\begin{figure}[!t]
  \centering
  \subfloat[Original image]{\includegraphics[width= 2in]{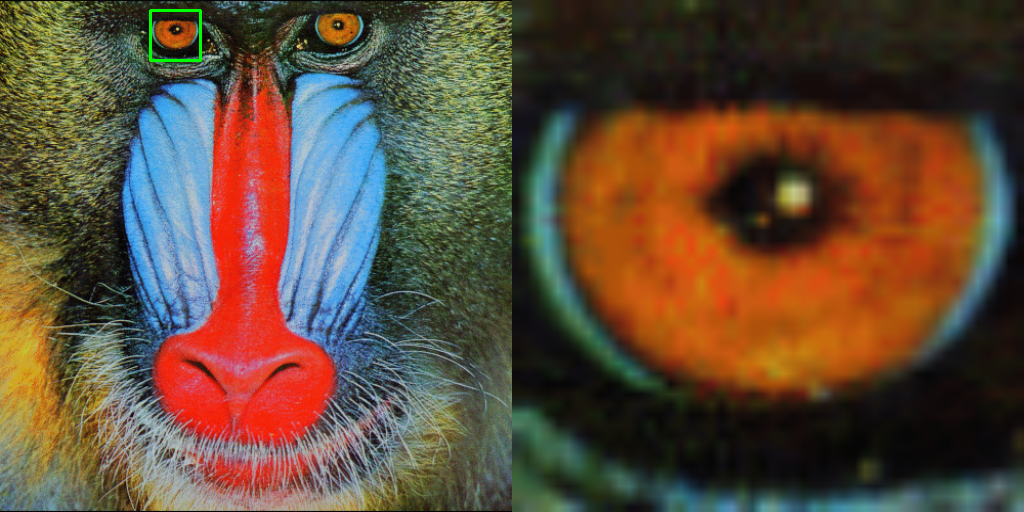}}\quad
  \subfloat[Noisy image]{\includegraphics[width= 2in]{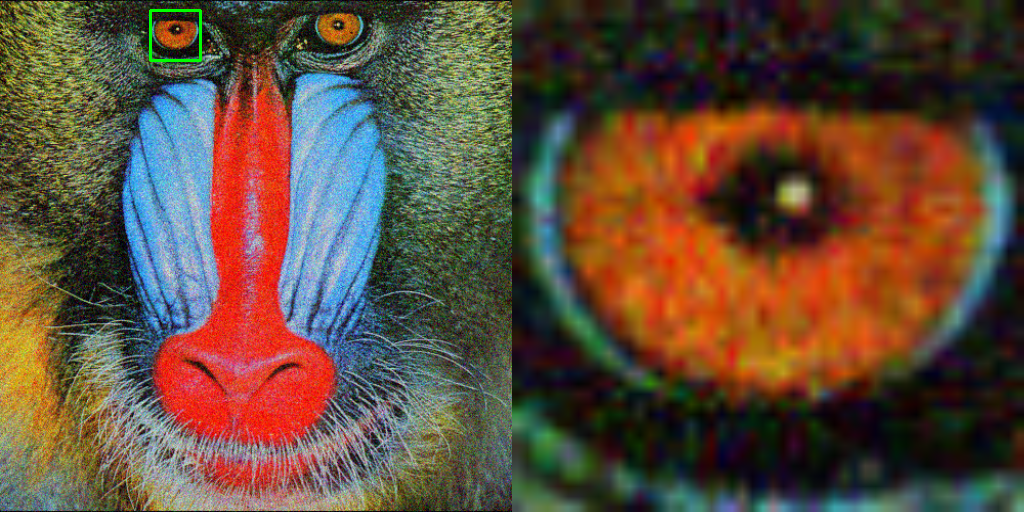}}\quad
  \subfloat[CBM3D]{\includegraphics[width= 2in]{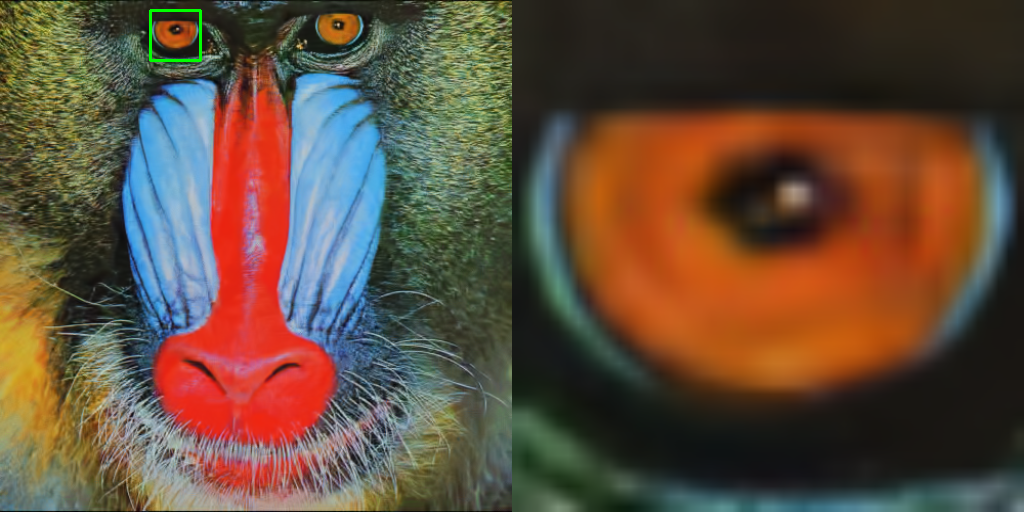}}\quad
  \subfloat[FFDNet]{\includegraphics[width= 2in]{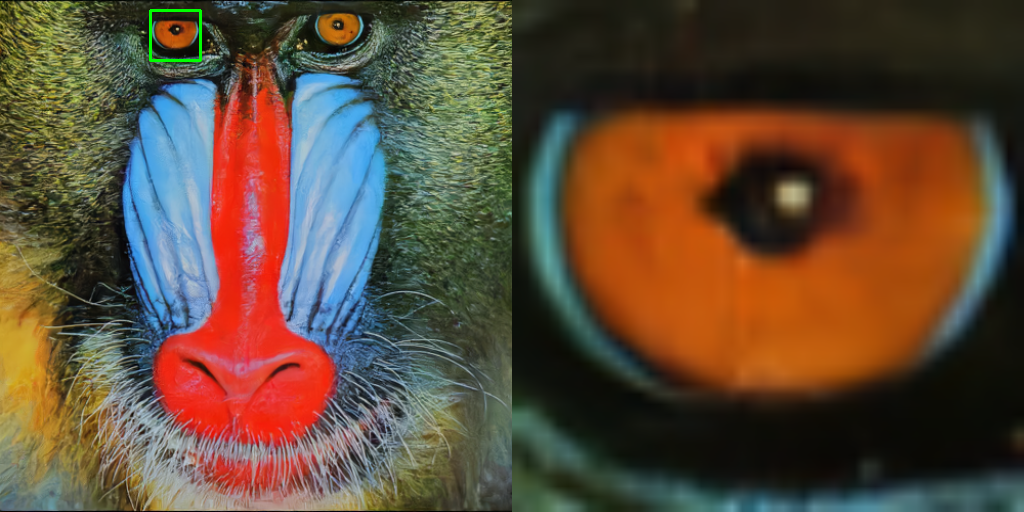}}\quad
  \subfloat[NFCNN]{\includegraphics[width= 2in]{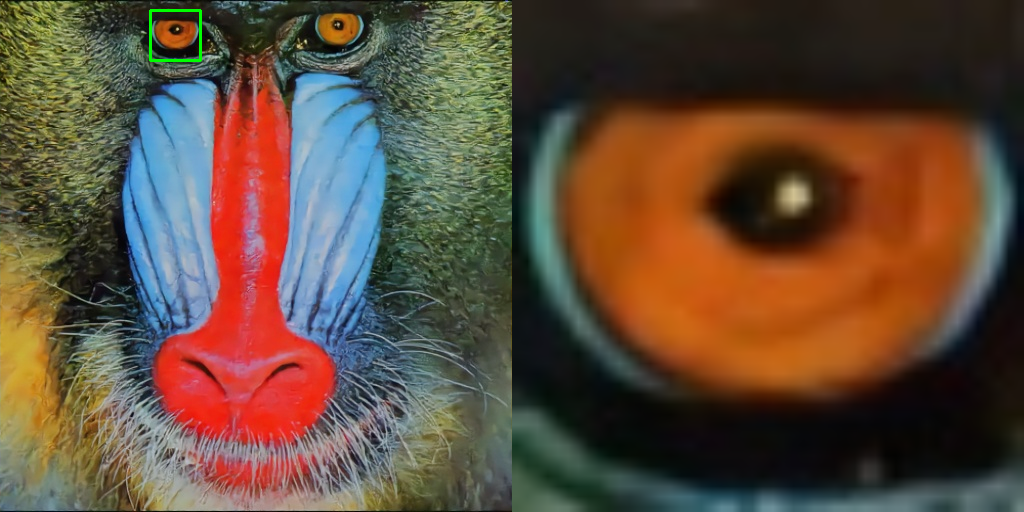}}\\
  \caption{Baboon with noise level $\delta=25$.}\label{baboon_delta25_compare}
\end{figure}

\begin{table}[!t]
  \caption{The average PSNR(dB) of different methods on BSD68 dataset with  noise level $\delta$  for gray scale images.}
  \label{BSD68_exp}
  \centering
  \scalebox{0.75}{
  \begin{tabular}{|c|c|c|c|c|c|c|c|c|}
  \hline {Methods} &{BM3D} &{WNNM} &{MLP} &{TNRD} &{DnCNN} &{FFDNet} &{NFCNN} \\
  \hline
  {$\delta=15$} & {31.07} & {31.37} & {-} & {31.42} & {31.72} & {31.63} & {\textbf{31.79}} \\
  \hline
  {$\delta=25$} & {28.57} & {28.83} & {28.96} & {28.92} & {29.23} & {29.19} & {\textbf{29.25}} \\
  \hline
  {$\delta=50$} & {25.62} & {25.87} & {26.03} & {25.97} & {\textbf{26.63}} & {26.29} & {\textbf{26.63}} \\
  \hline
  {$\delta=75$} & {24.21} & {24.40} & {24.59} & {-} & {24.64} & {24.79} & {\textbf{24.81}} \\
  \hline
  \end{tabular}
  }
\end{table}

\begin{table}[!t]
  \caption{The average PSNR(dB) of different methods on CBSD68 dataset with noise level  $\delta$ for color scale images.}
  \label{CBSD68_exp}
  \centering
  \scalebox{0.75}{
  \begin{tabular}{|c|c|c|c|c|c|c|c|c|}
  \hline {Methods} &{BM3D} &{DnCNN} &{FFDNet} &{NFCNN} \\
  \hline
  {$\delta=15$} & {33.52} & {33.89} & {33.87} & {\textbf{33.91}} \\
  \hline
  {$\delta=25$} & {30.71} & {31.23} & {31.21} & {\textbf{31.24}} \\
  \hline
  {$\delta=50$} & {27.38} & {27.92} & {\textbf{27.96}} & {27.94} \\
  \hline
  {$\delta=75$} & {25.74} & {24.47} & {26.24} & {\textbf{26.41}} \\
  \hline
  \end{tabular}
  }
\end{table}

\begin{table}[!t]
  \caption{The average PSNR(dB) of different methods on Set12 dataset with noise level $\delta$ for gray scale images.}
  \label{Set12_exp}
  \centering
  \scalebox{0.75}{
  \begin{tabular}{|c|c|c|c|c|c|c|c|c|}
  \hline {Methods} &{BM3D} &{WNNM} &{MLP} &{TNRD} &{DnCNN} &{FFDNet} &{NFCNN} \\
  \hline
  {$\delta=15$} & {32.37} & {32.70} & {-} & {32.50} & {32.86} & {32.75} & {\textbf{32.88}} \\
  \hline
  {$\delta=25$} & {29.97} & {30.26} & {30.03} & {30.06} & {\textbf{30.43}} & {\textbf{30.43}} & {30.42} \\
  \hline
  {$\delta=50$} & {26.72} & {27.05} & {26.78} & {26.81} & {27.18} & {27.32} & {\textbf{27.36}} \\
  \hline
  {$\delta=75$} & {24.91} & {25.23} & {25.07} & {-} & {25.20} & {25.49} & {\textbf{25.52}} \\
  \hline
  \end{tabular}
  }
\end{table}

\begin{table}[!t]
  \caption{The average PSNR(dB) of different methods on Kodak24 dataset with noise level $\delta$ for color scale images.}
  \label{CKodak24_exp}
  \centering
  \scalebox{0.75}{
  \begin{tabular}{|c|c|c|c|c|c|c|c|c|}
  \hline {Methods} &{BM3D} &{DnCNN} &{FFDNet} &{NFCNN} \\
  \hline
  {$\delta=15$} & {34.28} & {34.48} & {34.63} & {\textbf{34.72}} \\
  \hline
  {$\delta=25$} & {31.68} & {32.03} & {32.13} & {\textbf{32.15}} \\
  \hline
  {$\delta=50$} & {28.46} & {28.85} & {\textbf{28.98}} & {28.91} \\
  \hline
  {$\delta=75$} & {26.82} & {25.04} & {27.27} & {\textbf{27.29}} \\
  \hline
  \end{tabular}
  }
\end{table}

\begin{table}[!t]
  \caption{The average PSNR(dB) of different methods on McMaster dataset with noise level $\delta$ for color scale images.}
  \label{CMcMaster_exp}
  \centering
  \scalebox{0.75}{
  \begin{tabular}{|c|c|c|c|c|c|c|c|c|}
  \hline {Methods} &{BM3D} &{DnCNN} &{FFDNet} &{NFCNN} \\
  \hline
  {$\delta=15$} & {34.06} & {33.44} & {34.66} & {\textbf{34.71}} \\
  \hline
  {$\delta=25$} & {31.66} & {31.51} & {32.35} & {\textbf{32.43}} \\
  \hline
  {$\delta=50$} & {28.51} & {28.61} & {\textbf{29.18}} & {\textbf{29.18}} \\
  \hline
  {$\delta=75$} & {26.79} & {25.10} & {27.33} & {\textbf{27.36}} \\
  \hline
  \end{tabular}
  }
\end{table}

To demonstrate the effectiveness of NFCNN, some denoised results are displayed in \cref{delta_25_results}. We note that the textures of objects in pictures are still sharp after denoising. For example, \cref{baboon_delta25_compare} helps verify the texture preserving performance of NFCNN. The eye of baboon is cropped and resized to make a detailed comparison with different methods. The pictures are corrupted by AGWN with $\delta=25$. We can see that CBM3D leads to a over-smoothed result; FFDNet has  better denoising performance than CBM3D but fails to preserve the circle around iris, while NFCNN keeps a good balance between denoising and texture preserving. Notice that the artificial effect happens in the result of FFDNet, but NFCNN generates a more natural result. The pleasant texture preserving performance of NFCNN mostly benefits from the information exchange by the fusion block, which helps the model extract the details from the residual image.

\section{Conclusion}\label{section5}
We have proposed a deep learning based denoising method, called NFCNN, with a module of fusion block and a multi-stage structure. Thanks to the information exchange by the fusion block, NFCNN has a pleasant texture preserving ability. A stage-wise training strategy has been adopted in NFCNN to avoid the vanishing gradient and exploding gradient problems. Experimental results have verified the effectiveness of NFCNN, and demonstrated the competitive denoising performance when compared with the state-of-the-art algorithms.

\bibliographystyle{plain}
\bibliography{reference}
\end{document}